\documentclass[
 reprint,
 amsmath,amssymb,
 aps,
]{revtex4-2}

\usepackage{graphicx}
\usepackage{cancel}
\usepackage{dcolumn}
\usepackage{bm}
\usepackage{textcomp}
\usepackage{afterpage}
\usepackage{float}
\usepackage{placeins}
\renewcommand{\vec}[1][\mathbf]{#1}

\begin{document}

\preprint{APS/123-QED}

\title{Transversely pumped laser driven particle accelerator}

\author{Tanner Nutting, Nicholas Ernst, Alexander G. R. Thomas, and Karl Krushelnick}
\affiliation{G\'{e}rard Mourou Center for Ultrafast Optical Science, University of Michigan, 2200 Bonisteel Boulevard, Ann Arbor, Michigan 48109, USA}


\date{\today}

\begin{abstract}
We present a new acceleration scheme capable of accelerating electrons and ions in an underdense plasma. Transversely Pumped Acceleration (TPA) uses multiple arrays of counter-propagating laser beamlets that focus onto a central acceleration axis. Tuning the injection timing and the spacing between the adjacent beamlets allows for precise control over the position and velocity of the intersection point of the counter-propagating beam arrays, resulting in an accelerating structure that propagates orthogonal to the direction of laser propagation. We present the theory that sets the injection timing of the incoming pulses to accelerate electrons and ions with a tunable phase velocity plasma wave. Simulation results are also presented which demonstrate 1.2 GeV proton beams accelerated in 3.6 mm of plasma and electron acceleration gradients on the order of 1 TeV/m in a scheme that circumvents dephasing. This work has potential applications as a compact accelerator for medical physics and high energy physics colliders.
\end{abstract}

\maketitle


\section{Introduction}
\vspace{-5mm}
The concept of using plasma waves produced by lasers to accelerate electrons in an underdense plasma \cite{tajima1979laser} has matured into a well-developed field with impressive accomplishments such as experimental observations of electron beams on the order of 10 GeV \cite{gonsalves2019petawatt, aniculaesei2024acceleration} in recent years. Spatiotemporal control of a laser’s focal position and velocity \cite{froula2018spatiotemporal} has been proposed as a method to go beyond limits to acceleration in laser driven particle accelerators. Advanced focusing concepts offering spatiotemporal control of a laser’s focal position and velocity \cite{froula2019flying,miller2023dephasingless, pierce2023arbitrarily} have been proposed for a variety of applications \cite{palastro2020dephasingless, gong2024laser}. 
\\
\indent An application of particular interest for high energy physics \cite{hinchliffe2004tev} involves the use of spatiotemporally controlled lasers to produce a plasma wave with a phase velocity equal to the speed of light in vacuum, \emph{c}. This allows for electron acceleration at higher densities, enabling a larger accelerating field (since the accelerating fields scale as $E \propto n^{1/2}$). This has been investigated in simulations using pulse front tilt \cite{debus2019circumventing} and stepped echelons \cite{palastro2020dephasingless} for electron acceleration and chirped pulses for ion acceleration \cite{gong2024laser}.
\\
\indent An additional application of spatiotemporally structured light in underdense plasmas is the acceleration of ions which could be performed by matching the phase velocity of the plasma wave to the velocity of the accelerating ion bunch. Simulations of this concept have been demonstrated by the use of a chirped pulse with a diffraction grating and a lens such that the focal position of the laser moves transversely to the propagation of the laser \cite{gong2024laser}.
\\
\indent In this work, we present the concept of Transversely Pumped Acceleration (TPA), a versatile accelerator concept that allows for customized tuning of plasma wave phase velocity. The concept of transversely pumped laser acceleration naturally invites the use of arrays of high repetition rate laser beams and the use of efficient optimization techniques and machine learning. We demonstrate applications of this proof of principle concept using the OSIRIS Particle in Cell (PIC) code.

\begin{figure}[hbt!]
\includegraphics[width=\columnwidth]{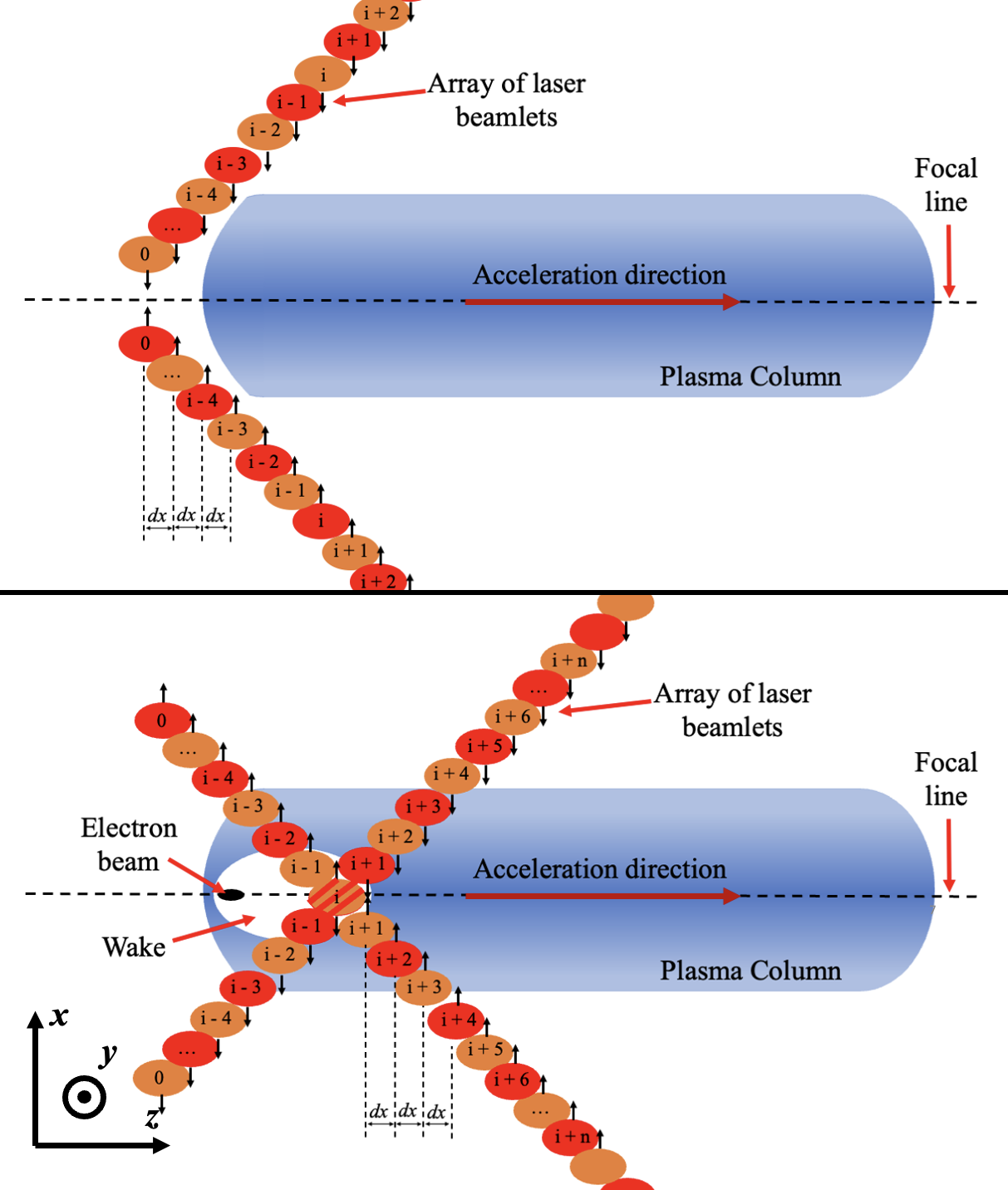}
\caption{\label{Fig:concept} Schematic of Transversely Pumped Acceleration scheme in the context of electron acceleration. Multiple laser beamlet arrays are injected symmetrically onto a line focus in the center of the plasma column. The direction of particle acceleration is transverse to the direction of laser beamlet propagation. Laser propagation directions are denoted with black arrows.}
\end{figure}

\section{Transversely Pumped Acceleration Concept}

TPA uses multiple arrays of counter-propagating, focused laser beamlets. The adjacent pulses within each laser beamlet array are delayed with respect to one another such that the intersection point of the incoming beamlet arrays on the focal line propagates at a desired velocity. A schematic of the proposed acceleration concept is shown in Figure \ref{Fig:concept}.
\\
\indent TPA offers attractive advantages over existing schemes for dephasingless electron acceleration as well as ion acceleration. Existing dephasingless electron acceleration concepts are limited when scaling up to obtain electron energies in the TeV range. Using stepped echelons to impart radial pulse delay onto a laser pulse to be focused by an axiparabola is a way to ensure a constant wake phase velocity of \emph{c} \cite{palastro2020dephasingless}. However, in order to reach TeV electron energies, one must operate the accelerator over several meters, requiring on the order of 10 ns of radial pulse delay. Another scheme proposes the crossing of two counter-propagating laser pulses with pulse front tilts to produce an intersection point that travels at \emph{c} \cite{debus2019circumventing}. This concept is limited by the amount of energy that can be put into the two laser pulses, requiring large and expensive optics to reach high energies. TPA is not limited by the aforementioned factors and is very straightforward to scale up in length, and therefore, electron energy gain. In this scheme it is only necessary to add more moderate energy laser pulses to the arrays of laser beamlets to achieve more acceleration. An additional advantage of using many moderate to low energy laser pulses in the TPA scheme is the ability to operate at a much higher repetition rate. This is desirable for industrial and medical applications, and enables the use of machine learning to optimize parameters.  The use of many low energy laser beamlets also allows for the use of smaller optics, significantly reducing cost. In addition, it is straightforward to create a plasma wave of arbitrary phase velocity using this scheme.
\\
\indent The advantages of TPA are also applicable to ion acceleration in an underdense plasma. The energy gain of the scheme in \cite{gong2024laser} is limited by the pulse length of the chirped laser pulse. For longer pulses, more energy needs to be added to the laser to maintain the same peak intensity. Additionally, the scheme in \cite{gong2024laser} requires complicated optics that may be difficult to manufacture, whereas TPA's most complicated aspects are alignment and timing, both of which can be controlled electronically. 
\\
\indent TPA's transverse geometry also allows for flexibility in the injection of beams into the accelerating structure. In this work, we simulate a proton beam that could be produced by Radiation Pressure Acceleration (RPA) \cite{macchi2010radiation} and inject it into the accelerating structure (see section \textit{Ion Simulation Results} for more details). Note that the proton beam to be accelerated could also be directly produced in the plasma.
\\
\indent Coherence effects from the incoming lasers in TPA can cause laser beating which negatively affects the stability of the accelerating structure for both ion and electron acceleration. In order to mitigate these coherence effects, TPA can use alternating cross polarized beamlets. Figure \ref{Fig:concept} illustrates a schematic of the Transversely Pumped Acceleration concept in the context of electron acceleration. Arrays of laser pulses, consisting of individual beamlets, propagate orthogonal to the acceleration direction, as shown by the arrows. The $i^{th}$ set of pulses cross at the focal line, which is shared by each set of pulses. The timing and spacing between the adjacent pulses can be adjusted to give an arbitrary and variable phase velocity of the accelerating structure, such as \emph{c}. Note that the adjacent beamlets are cross polarized to avoid coherence effects, as shown by the red and orange colors in Figure \ref{Fig:concept}.

In practice, controlling the phase velocity of the plasma wave requires control over the injection timing of each set of laser pulses. 
Consider a  coordinate system with the $\hat{\vec{z}}$ being the propagation direction. A train of pulses, each labeled $i$, propagating in the direction $\hat{\vec{k}}$ with $\hat{\vec{k}}\cdot\hat{\vec{z}}=0$ is timed to make a controllable intensity maximum on axis. The position of the  intensity maximum is $s(t)$ with the resulting intensity pulse  designed to move with an axial  group velocity function $V = ds/dt$ and with axial intensity profile $I(\zeta,t)$, where $\zeta = z-s(t)$. The pulses are assumed to have a slowly varying complex envelope $A_i(\vec{x},t)$, $||(\nabla A_i)/A_i||\ll k_i$ and $||(\partial_tA_i)/A_i||\ll \omega_i$, such that the vector potential $\vec{A}(\vec{x},t)$ of the summed combinations of the pulses may be described by
\begin{equation}
\vec{A}(\vec{x},t) =\Re\sum_i \vec{\hat{e}}_i A_i e^{i(\vec{k}_i\cdot\vec{x}-\omega_i t)} \;,
\end{equation}
where $\vec{k}_i$, $\omega_i$ are the (central) wavevector and frequency of each pulse and $\vec{\hat{e}}_i$ is a  vector describing the polarization, which is assumed to be a constant and normalized $|\vec{\hat{e}}_i|^2=1$. The ponderomotive force depends on $ \langle\vec{A}^2\rangle$, where the angle-brackets refer to a time average over fast oscillations. 
Therefore,
 \begin{eqnarray}
\langle \vec{A}^2\rangle &=&\Bigg\langle\frac{1}{4}\sum_{i,j}(\vec{\hat{e}}_i\cdot\vec{\hat{e}}_j) A_i A_j e^{i(\vec{k}_i+\vec{k}_j)\cdot\vec{x}-i(\omega_i+\omega_j) t} \nonumber\\
&+& (\vec{\hat{e}}_i\cdot\vec{\hat{e}}_j^*) A_i A_j^* e^{i(\vec{k}_i-\vec{k}_j)\cdot\vec{x}-i(\omega_i-\omega_j) t}+c.c.\Bigg\rangle\;.
\end{eqnarray}
\\
\indent Considering that the only contributions come from states where $A_i$ and $A_j$ overlap in space/time, by alternating the polarization between orthogonal states with sufficiently spatiotemporally spaced pulses, as in this work, or in general by choosing different frequencies for  the individual pulses so that the time average $\langle e^{ i(\omega_i-\omega_j) t}\rangle$  is zero except for $i=j$, this reduces to the \emph{incoherent sum},
 \begin{equation}
  \langle \vec{A}^2\rangle =\frac{1}{2}\sum_i  |A_i(\vec{x},t) |^2\;.
\end{equation}
\\
\indent In general, the full spatio-temporal profile of the resulting intensity profile can be derived. For simplicity, here we  only consider the formation of the intensity maximum on axis (at $\vec{x}_\perp=\vec{0}$ for a single train of pulses and its dependence on distance propagated, and assume that the transverse distribution is symmetric, with pulses entering simultaneously from a number of angles, $N_\theta$, about the $z$ axis. We further assume that each pulse has an identical form, described by a single function $g(z,t)\in[0,1]$, 
with each pulse focused on axis at a position along the $z$-axis, $z_i$, and arriving at $\vec{x}_\perp=\vec{0}$ at time $t_i$ such that $A_i(\vec{x}_\perp=\vec{0},z,t)/\sqrt{2}=a_ig(z-z_i,t-t_i)$, where $a_i$ is some relative amplitude of each pulse, or
 \begin{equation}
  \langle \vec{A}^2\rangle(\vec{x}_\perp=\vec{0},z,t) =\sum_i  a_i^2 |g(z-z_i,t-t_i) |^2\;.
\end{equation}
Setting $z_i = s(t_i)$ such that $\zeta = z-s(t_i)$ at $t=t_i$, which means that the axial intensity  profile of the controllable intensity maximum is simply the transverse profile of each pulse in the $z$-direction, and using the properties of the  Dirac delta distribution  $\delta(x)$,
 \begin{equation}
U_P(\zeta,t) =  \langle \vec{A}^2\rangle(\vec{x}_\perp=\vec{0},z,t) = \int |g(\zeta,t^\prime) |^2 \rho(t-t^\prime)  dt^\prime\;,
\end{equation}
where $\rho(t)=\sum_i a_i^2\delta(t-t_i)$,  i.e., the convolution of the function $g(\zeta,t)$ with a (non-uniformly spaced) comb of $\delta$-distributions with amplitude $a_i^2$.

Here, we chose the spacing of the pulses in $z$ to be uniform, $z_{i+1}-z_i =\Delta z = {\rm constant}$. This means that we may write the time delay in terms of the inverse function of $s(t)$ as $t_i = s^{-1}(z_i)$, with $s(t)$ defined above. However, this means that, in general, the temporal spacing between the pulses, $t_i$, is non-uniform. To see this, we may take the time average of $I(\zeta,t)$ over an interval $T$,
 \begin{equation}
\langle U_P(\zeta,t)\rangle =  \frac{1}{T(t)}\int_{T(t)} U_P(\zeta, t^\prime)  dt^\prime\;,
\end{equation}
where $\int_{T(t)}$ indicates the integral is taken over an interval of width $T$ containing $t$ and allowed to vary in width with $t$. Hence,
 \begin{eqnarray}
\langle U_P(\zeta,t)\rangle &= & \int |g(\zeta,t^\prime) |^2 \left[\frac{1}{T(t)}\int_{T(t)} \rho(t^{\prime\prime}-t^\prime)dt^{\prime\prime}\right]dt^\prime \nonumber\\
&\equiv& \int |g(\zeta,t^\prime) |^2 \langle\rho(t-t^\prime)\rangle dt^\prime\;,
\end{eqnarray}
where 
\begin{equation}
\langle\rho(t)\rangle = \frac{1}{T(t)}\int_{T(t)} \rho(t^\prime)dt^\prime = \sum_i \frac{a_i^2}{T_i}\int_{T_i} \delta(t^\prime-t_i) dt\;.
\end{equation}
\begin{figure*}[htbp]
\begin{center}
\includegraphics[width=\textwidth]{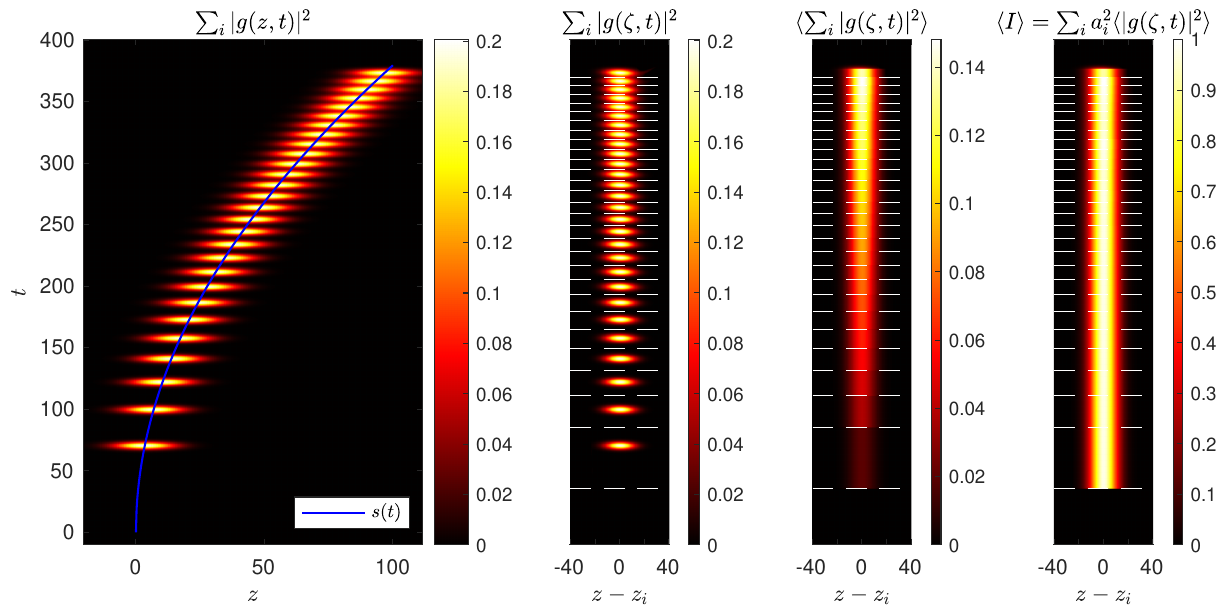}
\caption{Time averaging of pulse train. The colormap images show an example pulse train or time averaged pulse train as indicated in each subpanel title. The  blue line indicates the desired trajectory $s(t)$. The white dashed lines show the intervals for time averaging as described in the text. The final panel shows that a weighted intensity for each pulse leads to a constant time-averaged intensity peak. }
\label{timeav}
\end{center}
\end{figure*}
We now choose the interval $T_i$ so that it is of length $T_i = (t_{i+1}-t_{i-1})/2$, i.e., corresponding to a time  integration  over the range  $(t_{i}+t_{i-1})/2$ to $(t_{i+1}+t_{i})/2$. This means that the interval contains a single $\delta$-distribution corresponding to pulse $i$ and therefore $\langle\rho(t-t^\prime)\rangle$ is piecewise constant with magnitude $\frac{1}{T_i}$ in each interval between $\delta$-distributions with no gaps. Writing
\begin{equation}
T_i = \frac{t_{i+1}-t_{i-1}}{2} =  \frac{s^{-1}(z_{i+1})-s^{-1}(z_{i-1})}{2}\;,
\end{equation}
and noting that $s^{-1}(z) = t$ such that $ds^{-1}/dz = 1/V(z)$, where now the velocity function is expressed as a function of $z$ rather than $t$, we may use the Taylor expansion $s^{-1}(z_{i\pm1}) = s^{-1}(z_{i})\pm \Delta z ds^{-1}/dz + \dots$ to yield
\begin{equation}
T_i = \frac{\Delta z}{V_i} + \mathcal{O}(\Delta z^3 d^2/dz^2(1/V)|_i)\;,
\end{equation}
where $V_i = V(z_i)$. Taking only the leading order term under the assumption that $\Delta z^2 Vd^2(1/V)/dz^2|_i \ll 1$ and higher terms are smaller, we obtain
\begin{equation}
\langle\rho(t)\rangle =  \sum_i a_i^2\frac{V_i}{\Delta z}{\rm rect}\left[\frac{V_i(t-t_i)}{\Delta z}\right]\;,
\end{equation}
which is the aforementioned piecewise constant with magnitude $\frac{1}{T_i}$. Hence, if an effective constant ``density'' $\langle\rho(t)\rangle$ is required, then the relative amplitude of each pulse needs to be modulated by $a_i^2 \propto \Delta z / V_i$. However, if  $a_i^2 = \Delta z / V_i$ then $\langle\rho(t)\rangle = 1$ and therefore the peak intensity is at $\zeta=0$;
 \begin{eqnarray}
U_0 = \langle U_P(0,t)\rangle &= & \int |g(0,t^\prime) |^2dt^{\prime}\;,
\end{eqnarray}
Hence, the amplitude is required to be
\begin{equation}
a_i^2 = U_0\frac{\Delta z}{V_i\int |g(0,t^\prime) |^2dt^{\prime}} \;.
\end{equation}
Finally, for $N_\theta$ pulse trains crossing at the axis, we should divide the amplitude of each pulse by $N_\theta$ to achieve a ponderomotive potential of $U_0$. Hence, for a Gaussian pulse of the form $g(z,t) = \exp\left[-2{z^2}/{w_0^2}-2{t^2}/{\tau_0^2}\right]$, 
\begin{equation}
a_i^2 = U_0\sqrt{\frac{2}{\pi}}\frac{\Delta z}{N_\theta V_i\tau_0} \;.
\end{equation}
Therefore, for a (time averaged) intensity profile of the form $I = |I_0 h(\zeta)|$, where $h(\zeta)\in[0,1]$ describes the effective pulse shape, with $\zeta = z-s(t)$ with $s(t) = \int_0^t V(t^\prime) dt^\prime$, such that the effective pulse is traveling at group velocity $V$, then we need to combine a train of transverse pulses  described by the pulse shape
\begin{equation}
I_i = I_0\sqrt{\frac{2}{\pi}}\frac{\Delta z}{N_\theta V_i\tau_0} |g(z-z_i,t-t_i)|^2\;,
\end{equation}
with $|g(z-z_i,0)|^2 = h(\zeta)$,  $z_i = i\Delta z$ and $t_i = s^{-1}(z_i)$, with $s^{-1}(z) = \int_0^z dz^\prime/V(z^\prime)$ and where $I_i$ is the intensity of individual pulses in the train. The time averaging and weighting with $a_i^2$ to achieve constant time averaged intensity for a series of gaussian pulses with waist $w=4$ and duration $\tau=16$ is shown in Fig.~\ref{timeav}.

\subsection{Phase matching}
For a particle accelerated along the $z$ axis starting at $z=0$ in a wake with electric field, $E_z(z)$, from the equation of motion the differential of the particle momentum is  $dp_z = (qE_z/v_z)dz$. Hence, we may write
\begin{equation}
s^{-1}(z) = \int_0^{p_z(z)}  \frac{v_z}{V}\frac{dp_z^\prime}{qE_z}\;.
\end{equation}
Under phase matching conditions with $V(z)=v_z$ and with the (time averaged) driver being constant such that the electric field is (on time averaged) constant $E_z=E^*$, then 
\begin{equation}
s^{-1}(z) = \frac{p_z(z)-p_0}{qE^*}\;,
\end{equation}
where $p_0$ is the initial particle momentum and the $z$ dependent momentum $p_z(z) = mc\sqrt{\gamma^2-1}$ can be found by integrating the equation for the particle kinetic energy, $\gamma(z) mc^2 = \gamma_0mc^2+qE^*z$. Hence,
\begin{equation} \label{eq:18}
t_i = \frac{mc}{qE^*}\left(\sqrt{\left(\gamma_0+\frac{qE^*z_i}{mc^2}\right)^2-1}-\frac{p_0}{mc}\right)\;,
\end{equation}
Consider two cases of interest:
\begin{itemize}
\item Ultrarelativistic electron, $\left(\gamma_0+\frac{qE^*z_i}{mc^2}\right)^2\gg 1$ and $\frac{p_0}{mc}\simeq \gamma_0$,
\begin{equation}
t_i \simeq\frac{z_i}{c}\;,
\end{equation}
i.e., the group velocity is constant at the speed of light $c$.
\item Proton accelerated from rest, $\gamma_0=1$ and $p_0=0$, then
\begin{equation}
t_i = \frac{z_i}{c}\sqrt{1+\frac{2mc^2}{qE^*z_i}}\;.
\end{equation}
In the case that ${qE^*z_i}/{mc^2}\ll1$, i.e. the particle remains non-relativistic, then 
\begin{equation}
t_i \simeq \sqrt{\frac{2m z_i}{qE^*}}\;.
\end{equation}
\end{itemize}
In the following sections, we demonstrate these concepts in particle-in-cell simulations.

\section{Ion Acceleration Simulation Results}

\subsection{Tunable accelerating structure}

\begin{figure*}[t]
\includegraphics[width=\textwidth]{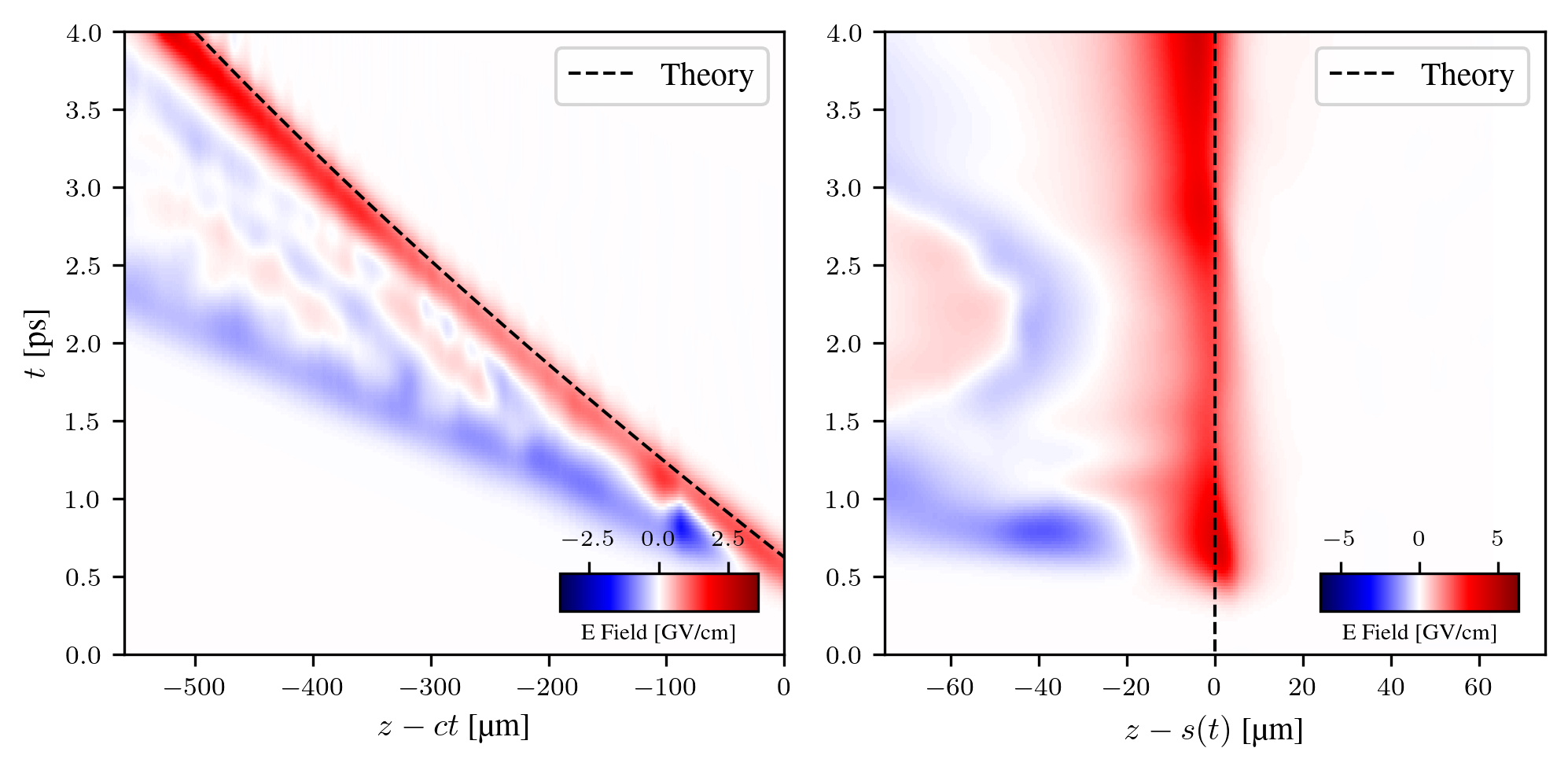}
\caption{\label{Fig:ion_streak} (Left) Streak plot of the on axis accelerating fields for ions in the speed of light frame. (Right) Streak plot of the same accelerating fields in the frame co-moving with the ion beam that is being accelerated. The theoretical predictions for both streak plots are shown as a dashed black line.}
\end{figure*}
\begin{figure}[h]
\includegraphics[width=\columnwidth]{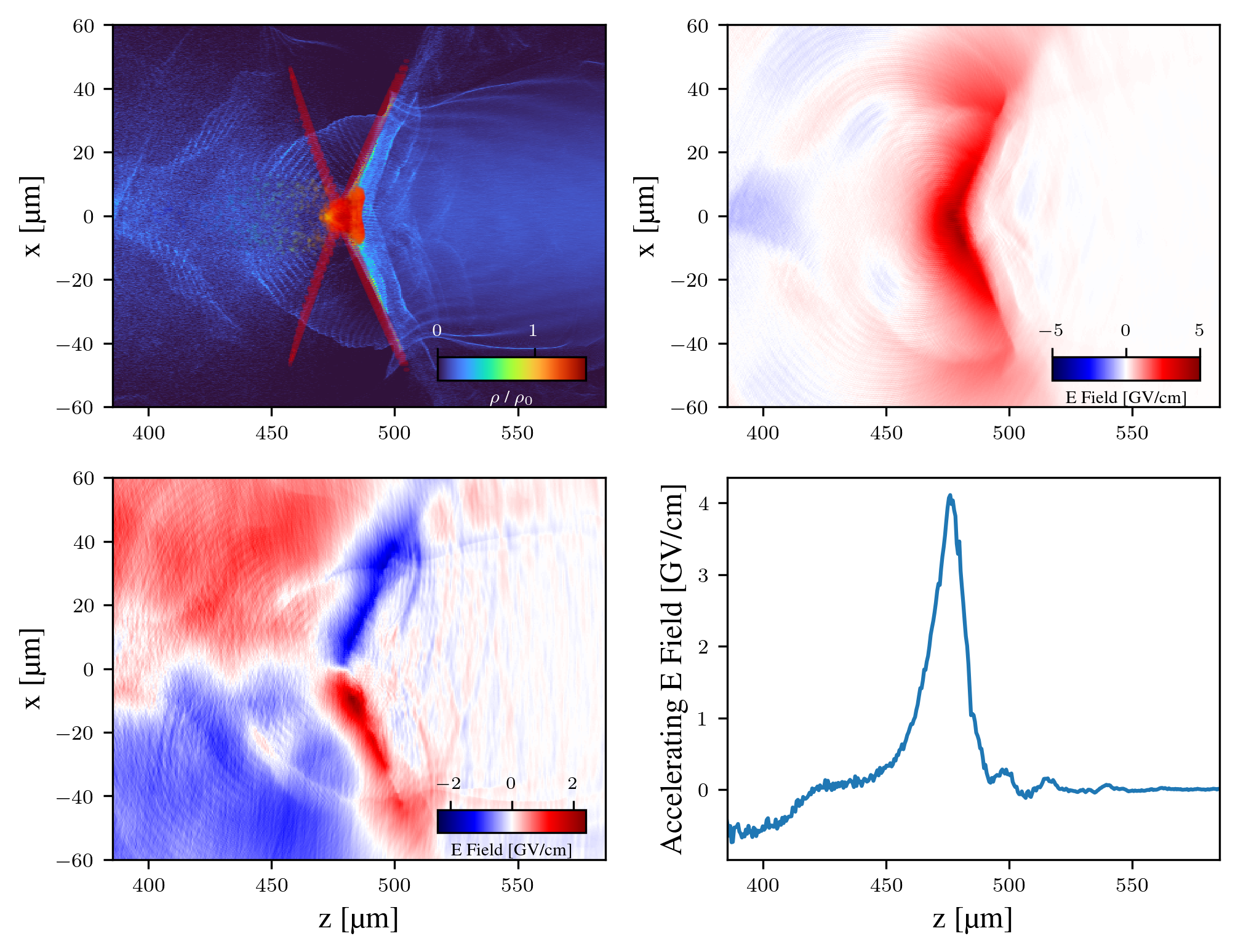}
\centering
\caption{\label{Fig:ion_beam_and_lasers} (Top left) Ion beam in red, shown in accelerating structure. The electron density is shown in blue and the laser pulses are shown as red contours. (Top Right) Accelerating field structure. (Bottom left) Focusing field structure. (Bottom right) Line-out of accelerating field structure along the central axis of the simulation.}
\end{figure}

A structure suitable for the acceleration of heavier, positively charged particles, such as protons, has been developed and tested using the theoretical tools developed in the previous section. For this study, consider a quasi-monoenergetic proton beam with peak energy at 95 MeV and a FWHM energy spread of 20 MeV, reasonably produced by an RPA solid target source \cite{robinson2008radiation}. Protons with an energy of 95 MeV have a velocity of approximately 0.42\emph{c}, and can be injected into the TPA accelerating structure. It is critical to match the initial phase velocity of the accelerating plasma wave to the initial velocity of the injected ion beam. This can be done by inserting the correct value of $\gamma_0$ into Eq. \ref{eq:18}.
\\
\indent To illustrate that the accelerating structure produced in simulations matches the theory of the previous section, two streak plots of the on axis accelerating fields produced by the plasma wave are shown in Figure \ref{Fig:ion_streak}. Both plots are of the same simulation, but are presented in different frames of reference. The left figure illustrates the temporal evolution of the accelerating fields in the speed of light frame. The accelerating structure starts slower than the speed of light (0.42\emph{c}) and accelerates in time. As one may expect, the accelerating region slips back in the speed of light frame, and follows a curve with a positive second derivative due to the acceleration of the structure. If run longer, the curve traced out by accelerating region would asymptotically approach a vertical line as the speed of the structure nears \emph{c}.
\\
\indent The right plot in Figure \ref{Fig:ion_streak} illustrates the accelerating field of the structure in the frame co-moving with the ion beam that is to be accelerated. In agreement with the theory, the accelerating region traces a vertical line. This indicates that the position of the accelerating region is nearly constant in the frame $z-s(t)$.
\\
\indent Figure \ref{Fig:ion_beam_and_lasers} illustrates an example of the accelerating structure for ions. An ion beam, shown in red in the top left figure, is accelerated along the leading edge of the crossing point between the incoming laser beamlets. TPA parameters can be tuned to provide a favorable accelerating structure for either positive or negative charges. By increasing the laser intensity relative to the density of the plasma, more electrons are expelled from the region near the crossing point of the beamlets. As this point propagates, a field structure that can accelerate positive ions is produced.
\\
\indent This accelerating structure is highly tunable. The initial velocity and the acceleration of the structure is set by the transverse spacings of the laser beamlets and the temporal delays between the pulses. The strength of the accelerating field depends both on the plasma density and the laser beamlet intensity, since a higher intensity results in the displacement of more electrons and therefore a higher electric field near the crosing point of the beamlets. Note that the structure produced in Figure \ref{Fig:ion_beam_and_lasers} was produced with adjacent beams in phase with one another and polarized in the $\hat{y}$ direction for the purpose of field visualization. All other results are shown for cross polarized adjacent beamlets.
\subsection{Acceleration of proton beams to relativistic energies}
\indent With an accelerating structure that ensures phase matching between the accelerating ion beam and the velocity of the plasma wave, the acceleration of protons to relativistic energies can also be demonstrated. The injected ion beam has a Gaussian density profile in both dimensions with FWHM = 4 \textmu m and has a density of $1.5 \times 10^{16} \text{cm}^{-3}$. If cylindrical symmetry were assumed for this beam, the resulting charge would be on the order of $\sim$ 10 pC.
\\
\indent The accelerating structure was produced in an underdense hydrogen plasma with density $1.8 \times 10^{18} \text{cm}^{-3}$. Each beamlet of the incoming arrays had a normalized vector potential $a_0 = 8$, beam waist $w_0 = 3$ \textmu m, pulse duration $\tau = 25$ fs, and spacing $dx = 3$ \textmu m. These parameters correspond to laser pulses with 290 mJ of laser energy per pulse. State of the art technology can produce such pulses at 1 kHz. 1200 beamlets were used per array to accelerate protons over a distance $1200 \times dx = 3.6$ mm. Adjacent laser pulses were cross polarized with respect to one another, polarizations are in $\hat{z}$ and $\hat{y}$ directions.
\begin{figure}[H]
\includegraphics[width=0.48\textwidth]{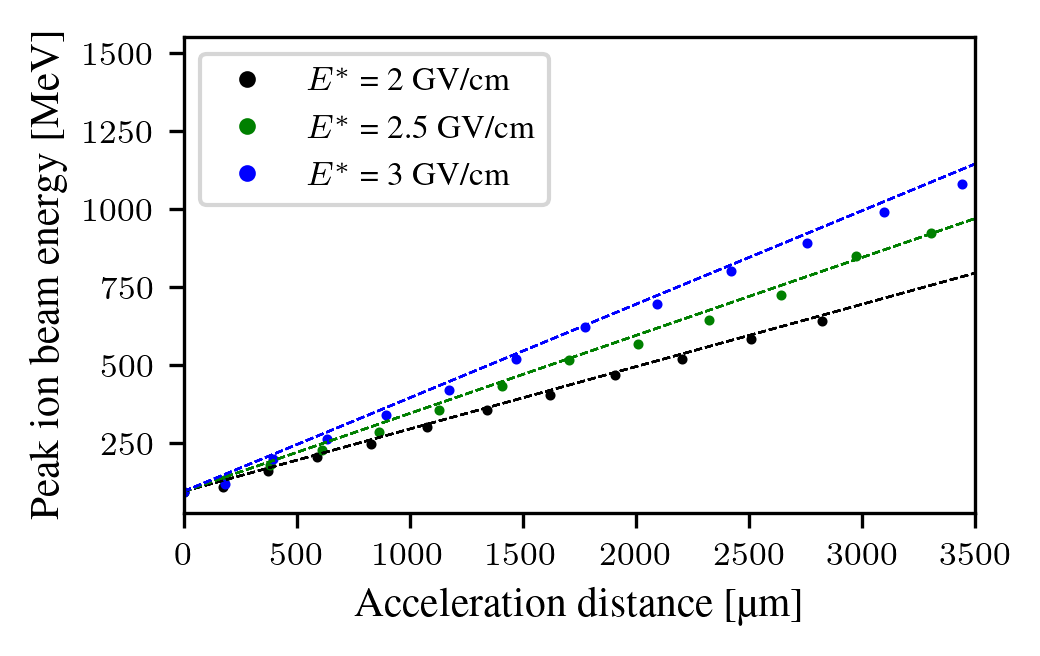}
\caption{\label{Fig:Enevdist} Peak ion energy as a function of acceleration distance for different values of $E^{*}$. A maximum proton energy of 1.2 GeV was obtained in 3.6 mm for $E^{*} = 3$ GV/cm.}
\end{figure}
\vspace{1 mm}
\indent A plot showing the proton energy gain vs. acceleration length for the simulation parameters described above for various values of $E^{*}$ is shown in Fig. \ref{Fig:Enevdist}. The dashed lines on the plot indicate theoretical predictions. Each simulation used the same plasma and laser parameters. The maximum ion energy obtained for a value of $E^{*} = 2.5$ GV/cm was 1.12 GeV, with a peak energy of 0.98 GeV, demonstrated after 3.6 mm of acceleration.
\begin{figure}[H]
\includegraphics[width=\columnwidth]{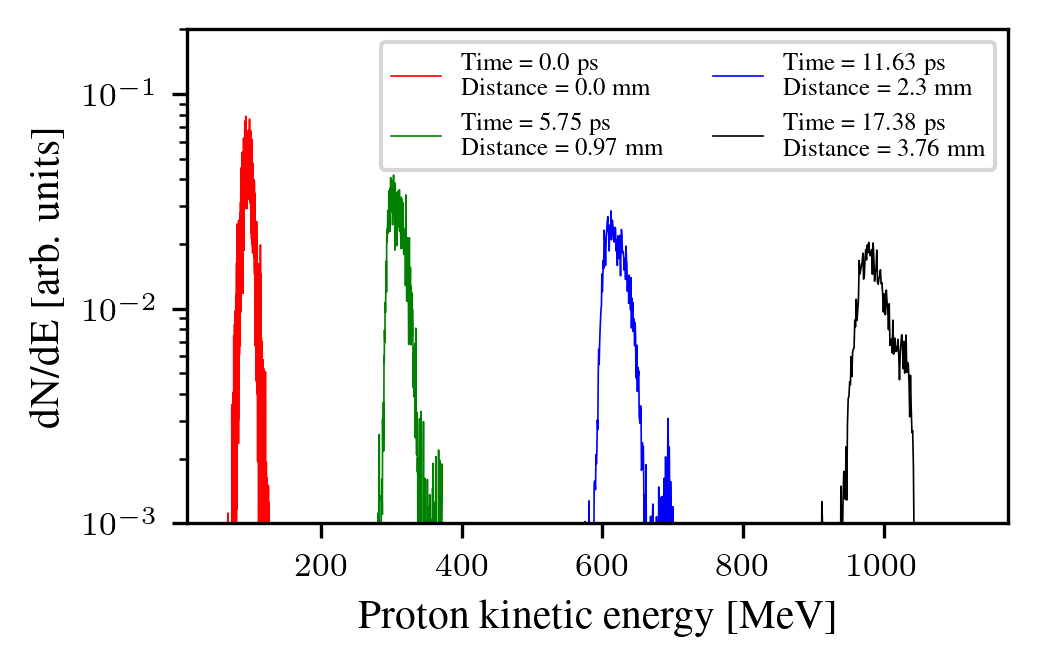}
\caption{\label{Fig:ispec} Ion energy spectrum at different times for $E^{*} = 2.5$ GV/cm.}
\end{figure}
\indent The quasi-monoenergetic nature of the initial ion beam is preserved during acceleration in the TPA scheme. Fig. \ref{Fig:ispec} shows the ion spectrum at 4 different times during the acceleration process for a value of $E^{*} = 2.5$ GV/cm. The energy spread of the initial injected ion beam is 20 MeV FWHM while the final energy spread is 40 MeV FWHM. This result demonstrates the stability of the structure and the effectiveness of the phase matching over several millimeters. Similar structures could in principle also be used for positron acceleration.
\section{Dephasingless Electron Acceleration Simulation Results}
\subsection{Accelerating structure}
\indent TPA's versatility enables its use for dephasingless electron acceleration as well. This is possible due to the ability to set an arbitrary phase velocity of the plasma wave. By setting the velocity of the accelerating structure equal to \emph{c}, dephasingless operation is achieved.  This allows for the use of higher density plasmas, therefore increasing acceleration gradients. The accelerating structure in the case of electron acceleration is shown in Figure \ref{Fig:acc_struct_electron}. The plots in this figure are produced with adjacent laser beamlets that are in phase with one another and polarized in the $\hat{y}$ direction. This was again done for the purpose of illustrating the focusing and accelerating fields. Unless explicitly mentioned otherwise, all of the results in this section were produced using cross polarized adjacent beamlets since this is more practical for potential implementation of this concept. 
\\
\indent Appropriate parameters for electron acceleration using the TPA concept are as follows: spacing between adjacent beamlets $dx = 1.3$ \textmu m, beam waist of laser $w_0 = 1.64$ \textmu m, pulse duration $\tau = 20$ fs, normalized vector potential $a_0 = 2.5$, laser wavelength $\lambda = 1$ \textmu m and plasma density $\rho = 1.2 \times 10^{18} \text{cm}^{-3}$ (or 0.012$n_{cr}$). These parameters correspond to laser pulses with energy equal to 7 mJ, readily produced at high repetition rate.
\setcounter{figure}{6}  
\addtocounter{figure}{0}  
\begin{figure}[H]
\includegraphics[width=0.47\textwidth]{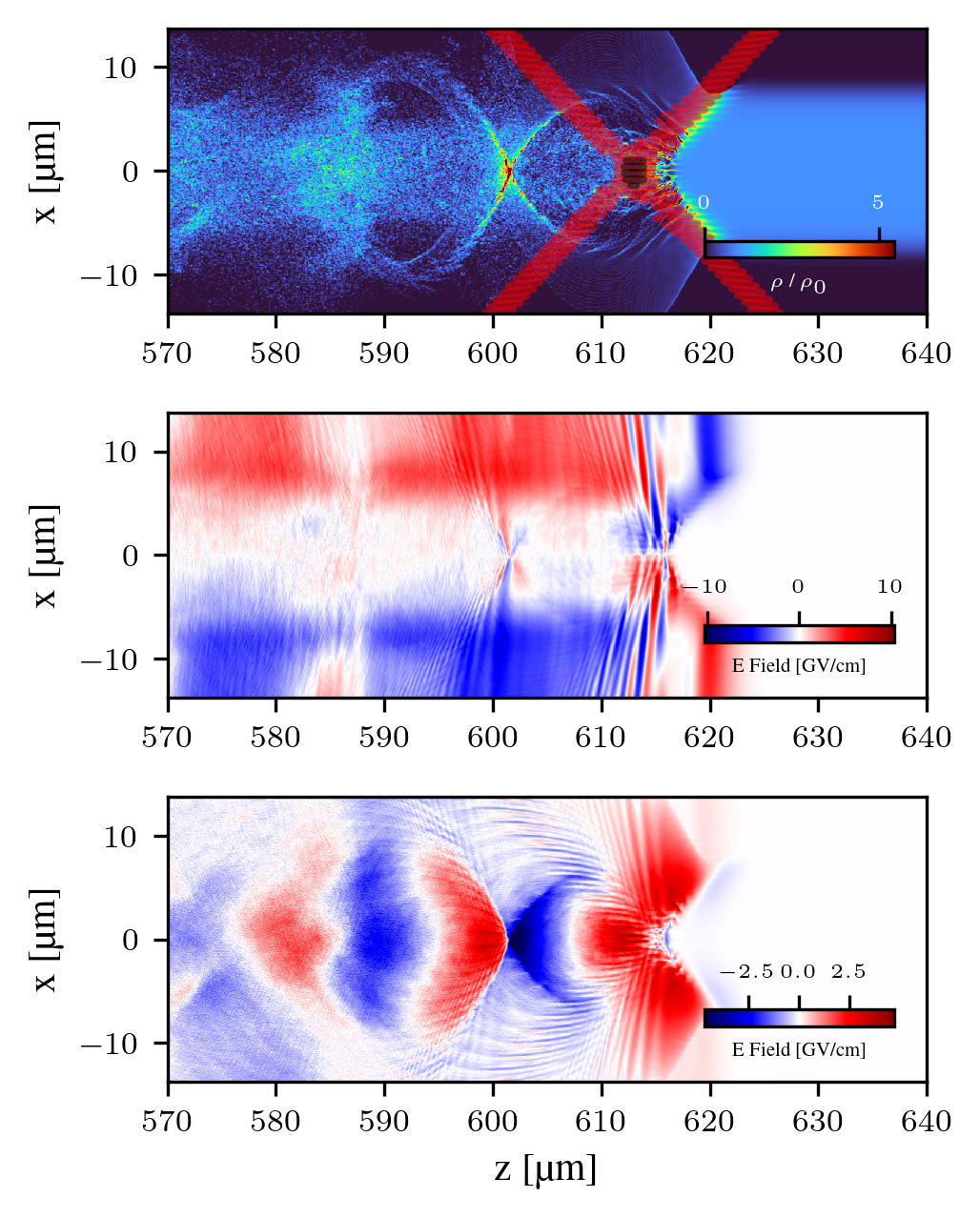}
\caption{\label{Fig:acc_struct_electron} (Top) Electron density of accelerating structure. Laser contours are shown in red. The energy per laser pulse is 7 mJ. (Middle) Focusing fields. (Bottom) Accelerating fields. Note that negative fields are shown in blue and positive fields are shown in red.}
\end{figure}
\indent Using higher densities will produce larger accelerating gradients while still avoiding dephasing. However, the intensity of each laser beamlet must be increased for higher density operation in order to produce the accelerating structure. For this study we chose to operate at $\rho = 1.2 \times 10^{18} \text{cm}^{-3}$ since this density was high enough to demonstrate the out-performance of a traditional LWFA in terms of avoiding dephasing. Yet, the chosen density was low enough to demonstrate what modest energy laser beamlets (7 mJ) are capable of producing in the TPA arrangement.
\\
\indent Note the difference in accelerating structure between the electron and ion accelerators. The blowout of the plasma at the intersection of the incoming beamlets is less pronounced for the electron accelerator by reducing laser intensity relative to the plasma density. This results in an accelerating wave that trails behind the intersection of the beamlet crossing point by approximately 10 \textmu m. This trailing electron plasma wave produces fields favorable for the acceleration of negative charges. 
\\
\begin{figure}[H]
\centering
\includegraphics[width=\columnwidth]{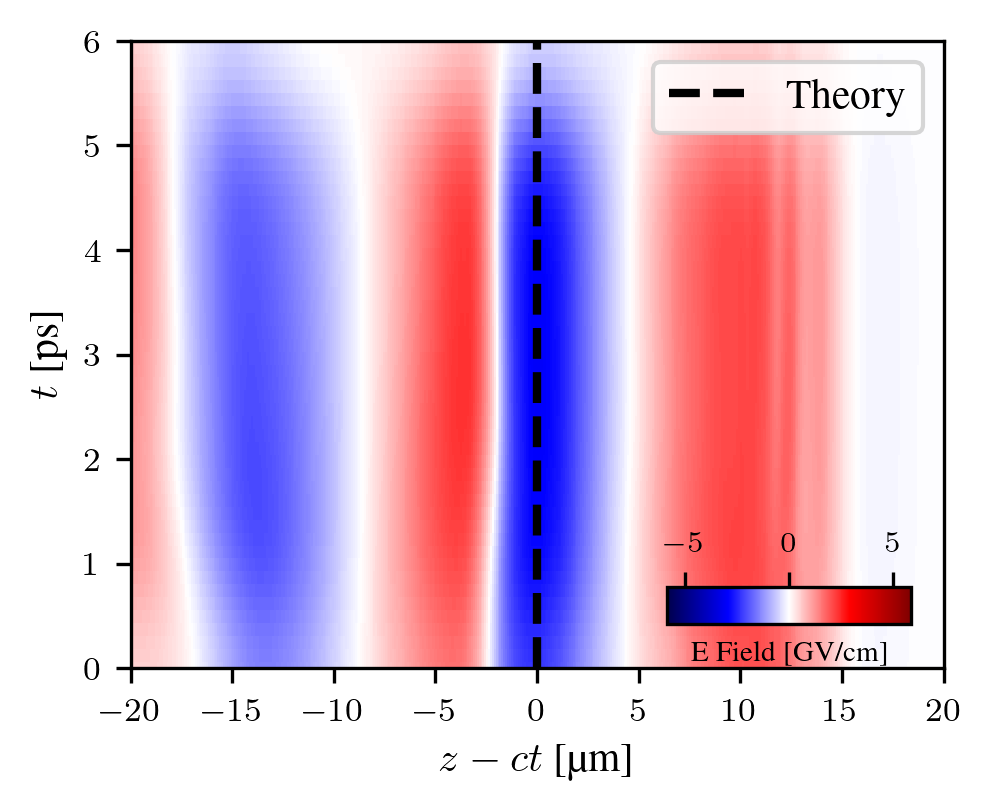}
\caption{\label{Fig:streak} Streak plot showing a phase velocity of \emph{c}. The on axis accelerating fields are shown as a function of time and space. The black dashed line denotes a perturbation with a velocity of \emph{c}, showing agreement of simulation with theory.}
\end{figure}
\indent In a similar fashion to that done for the ion acceleration case, the phase velocity of the accelerating structure is confirmed to be \emph{c} through a streak plot shown in Figure \ref{Fig:streak}. Here the on axis accelerating fields are plotted in the speed of light frame. The region where electrons are accelerated appears stationary in the speed of light frame, implying electrons being accelerated in the wake will never dephase. The vertical dashed line shows the theoretical prediction of the location of the accelerating structure in both time and space.
\subsection{Monoenergetic beam production and localized injection}
\begin{figure*}[t]
\includegraphics[width=\textwidth]{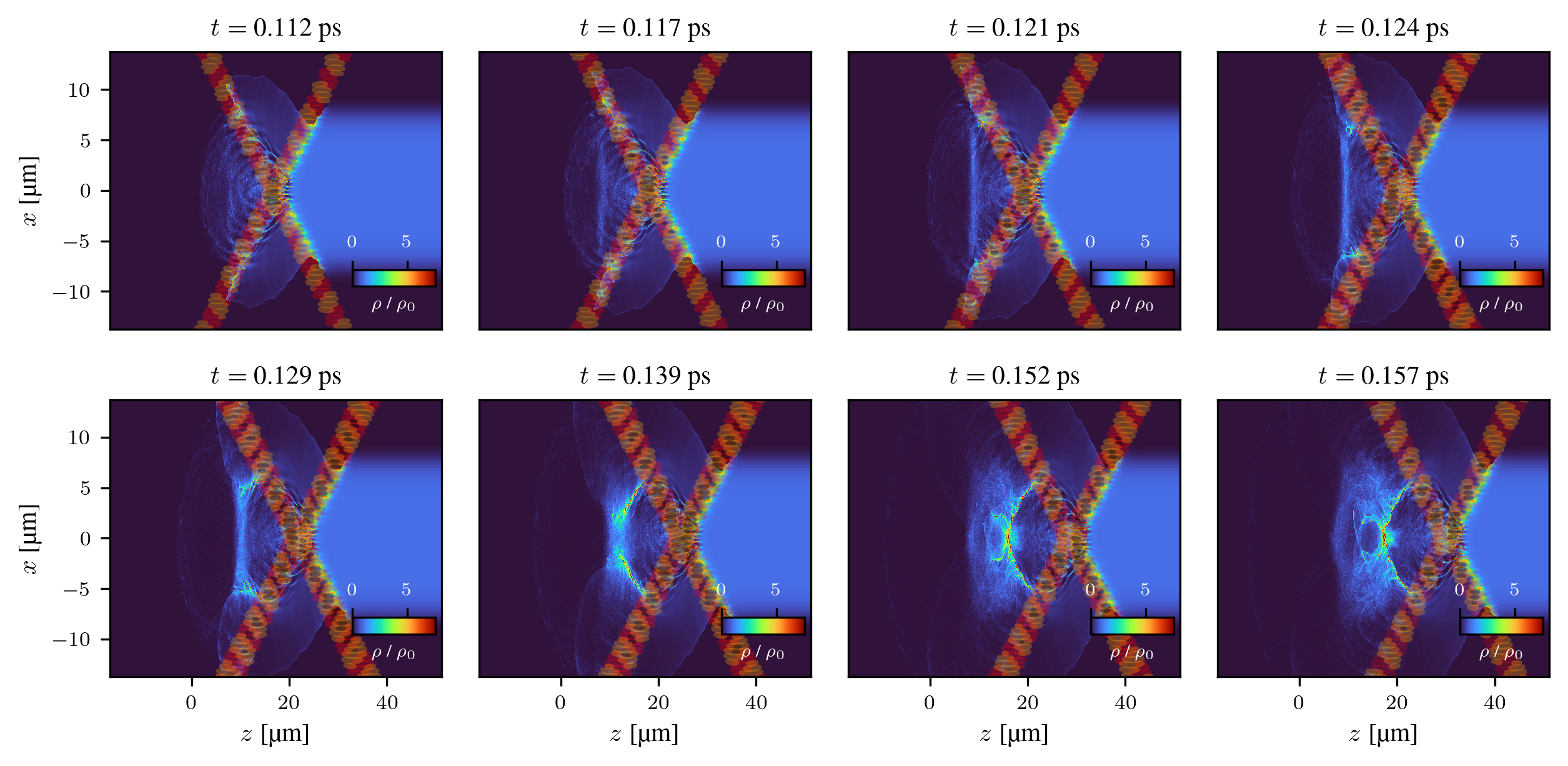}
\caption{\label{Fig:dynamics} Time evolution of the wake formation and injection dynamics. Earliest time is in the top left, latest time is in the bottom right.}
\end{figure*}
\begin{figure}[h]
\includegraphics[width=\columnwidth]{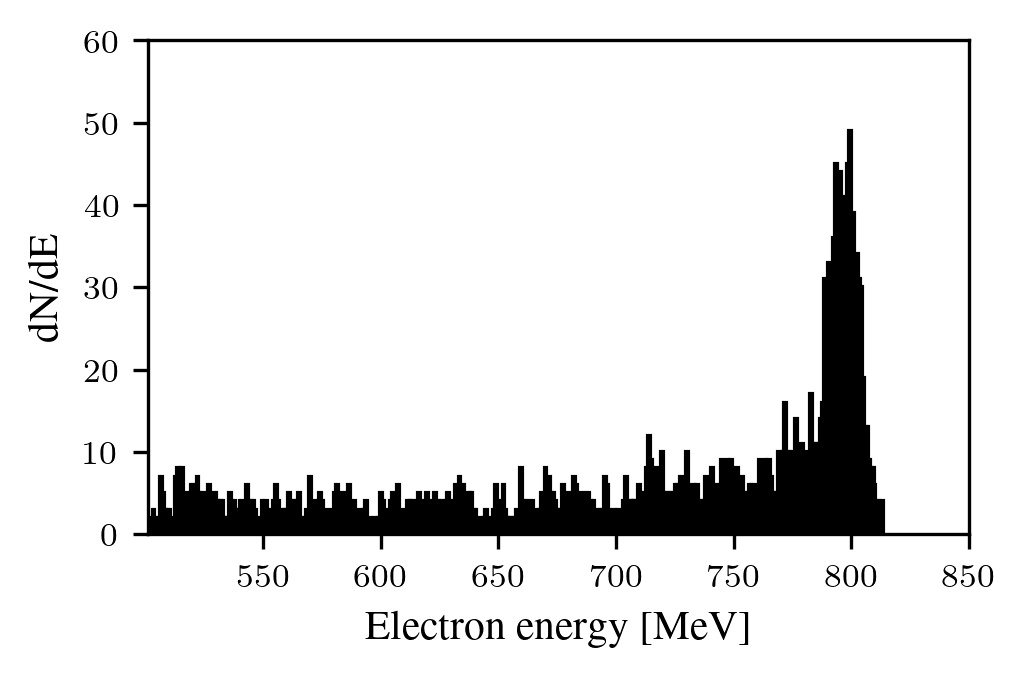}
\caption{\label{Fig:mono} Electron spectrum of monoenergetic beam. The energy spread is 1.9\%.}
\end{figure}
\begin{figure}[H]
\includegraphics[width=\columnwidth]{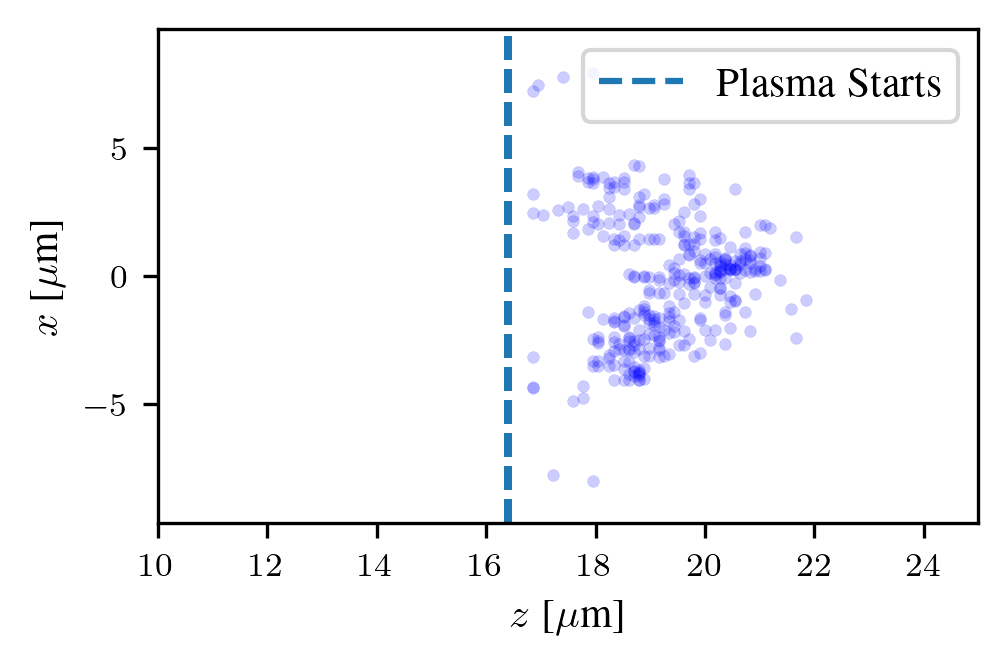}
\caption{\label{Fig:inital_parts} Initial positions of the electrons in the 800 MeV monoenergetic electron beam. Note that the particles are injected within 5 \textmu m of the start of the plasma.}
\end{figure}

\indent Using a similar accelerating structure to that discussed in the previous section, monoenergetic electron beams can be produced using TPA. By injecting the electron beam at the start of the plasma column and accelerating over 1525 \textmu m, an 800 MeV electron beam was produced with a 1.9\% energy spread, resulting in an acceleration gradient of 0.52 TeV/m. The emittance of the beam was calculated to be 1.38 mm-mrad. The spectrum of this beam is shown in Figure \ref{Fig:mono}.
\\
\indent Using the scalings in \cite{PhysRevSTAB.10.061301}, the dephasing length of a traditional laser wakefield accelerator operating at the same density of 0.012 $n_{cr}$ can be calculated to be 360~\textmu m. This calculation assumes matched conditions with a beam waist of 50 \textmu m and a laser pulse with the same total energy as our accelerating structure (16.4 J) in this specific example.
\\
\indent Producing a monoenergetic beam requires localized injection at the start of the plasma. Figure \ref{Fig:dynamics} shows plots of a time series of the injection event and wake formation. Notice that the electron charge gets pushed away from the center of the plasma column by the laser beamlets, until the attractive force from the ion channel that gets left behind overcomes the ponderomotive push of the beamlet arrays. Then, the electrons in the plasma collapse back into the plasma column, which produces a wake and injects charge into the accelerating structure at a specific location and time of the simulation. This structure is then continually sustained by additional incoming laser beamlets until a desired beam energy is produced.
\\
\indent By tracking the particles in the monoenergetic beam throughout the simulation, we can plot the initial positions of the particles in the beam. Figure \ref{Fig:inital_parts} illustrates the localized injection within 5 \textmu m of the start of the plasma column. This localized injection is a result of the sudden collapse of the plasma column back onto itself after the initial perturbation by the beamlet arrays.\\

\subsection{Three-dimensional acceleration results}
\begin{figure}[h]
\includegraphics[width=\columnwidth]{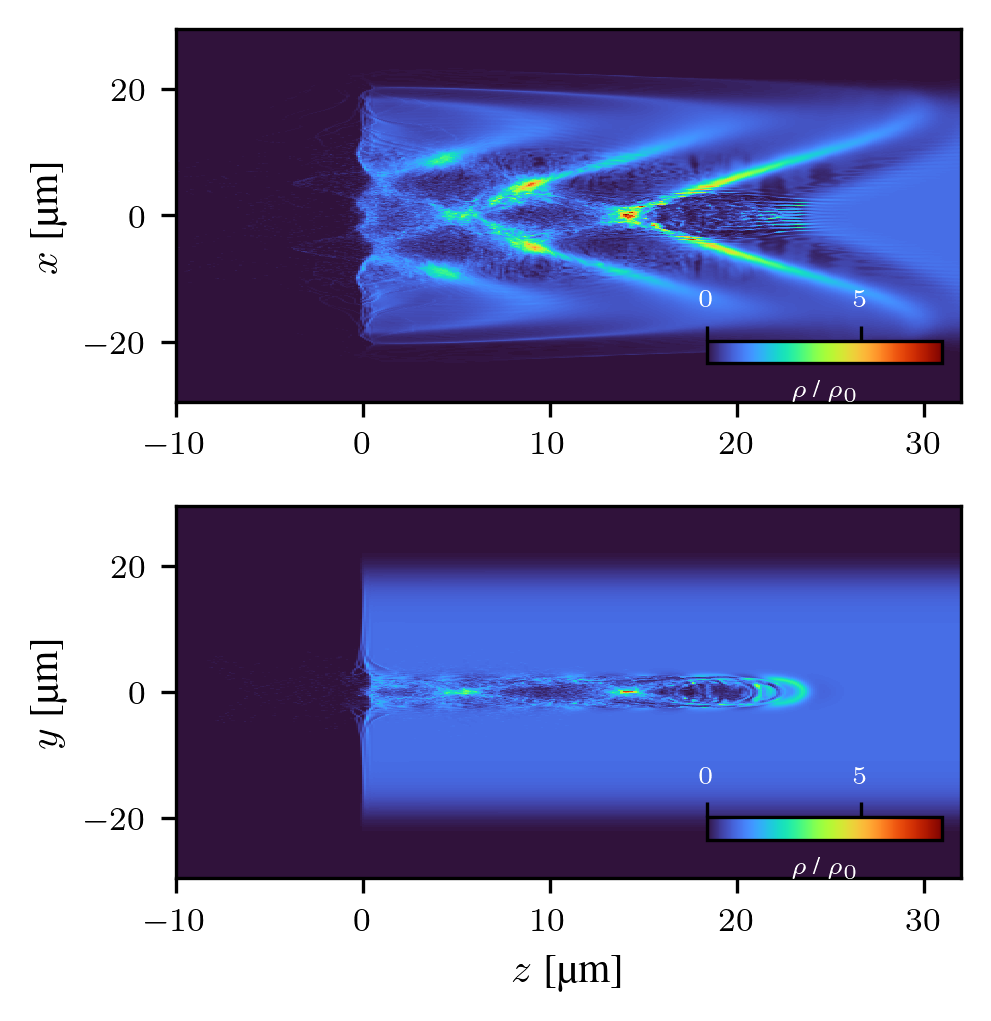}
\caption{\label{Fig:wide_3D} Cross sections of electron density profile for 3D simulation demonstrating the accelerating structure in a wider plasma. (Top) Cross section of the $y=0$ plane. (Bottom) Cross section of the $x=0$ plane.}
\end{figure}

Proof of principle simulations were conducted for the case of electron acceleration in three dimensions using OSIRIS. The conducted study examined the simplest case of $N_\theta = 2$, that is two beamlet arrays. Because of computational cost limitations, we only simulated these structures for short distances. These beamlet arrays were injected into the simulation box in the same plane ($y = 0$) and came to focus along a central axis, were $ x = y = 0$. The individual beamlets traveled in the $\hat{x}$ direction and crossed at $x = 0$, where the beams came to focus and the accelerating structure was produced. 
\\
\indent Two cases are presented here. The first of which uses the following simulation parameters: plasma column width $= 16$ µm FWHM, spacing between adjacent beamlets $dx = 1.3$ \textmu m, beam waist of laser $w_0 = 1.64$ \textmu m, pulse duration $\tau = 20$ fs, pulse energy $E = 7$ mJ, laser wavelength $\lambda = 1$ \textmu m and plasma density $\rho = 1.2 \times 10^{18} \text{cm}^{-3}$ (or 0.012$n_{cr}$). These parameters correspond to laser pulses with normalized vector potentials equal to $2.5$. 
\begin{figure}[h]
\centering
\includegraphics[width=\columnwidth]{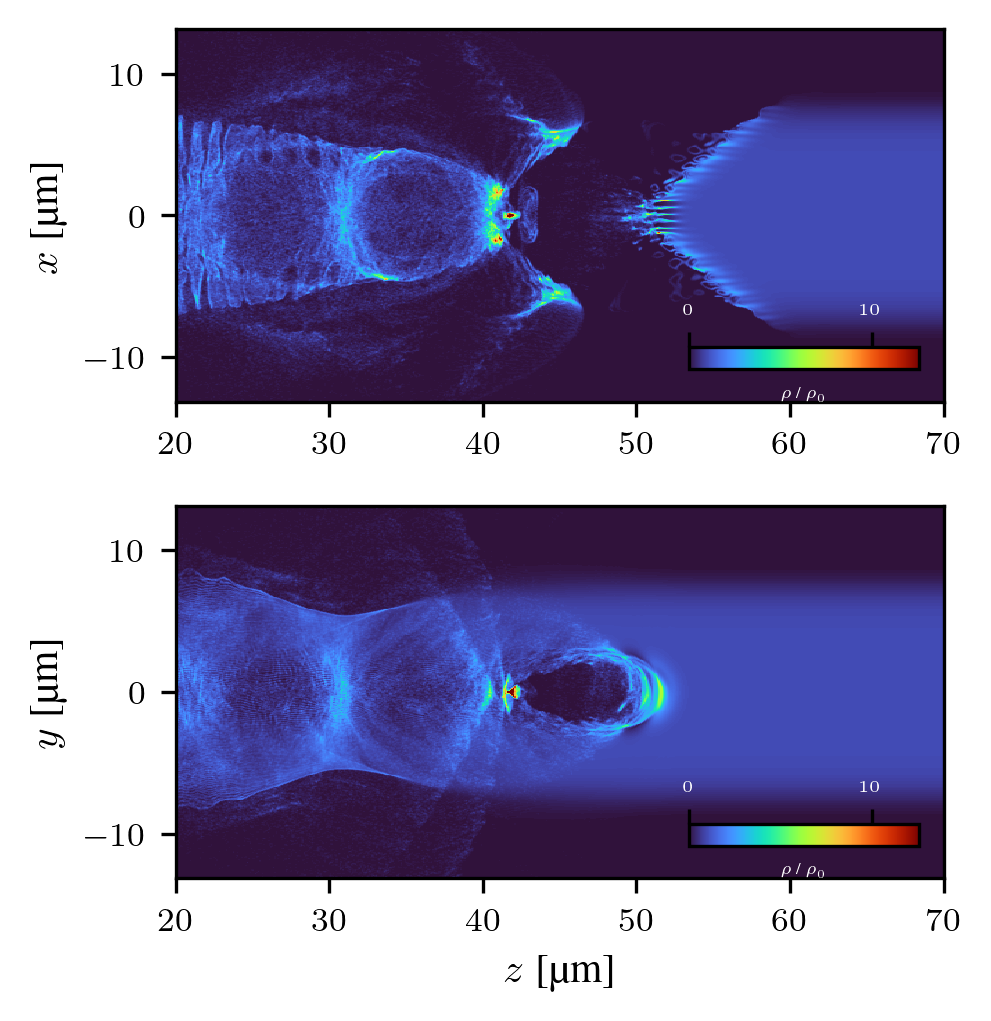}
\caption{\label{Fig:3D_2.5} Cross sections of electron density profile for 3D simulation. (Top) Cross section of the $y=0$ plane. Note that this is the plane that the laser pulses travel in. (Bottom) Cross section of the $x=0$ plane. The lasers travel in and out of the page in this view.}
\end{figure}
\\
\indent The accelerating structure formed in the electron density profile is shown in Figure \ref{Fig:3D_2.5}. The top and bottom plots are slices through the center of the plasma profile. The top slice shows the plane in which the lasers are traveling. The lasers are traveling from top to bottom and vice versa in the top slice. The lasers are traveling in and out of the page in the bottom slice. Note that the top slice is the $y=0$ plane and the bottom slice is the $x = 0$ plane.
\\
\indent The maximum energy in the electron beam produced in this simulation is 20 MeV after an acceleration distance of 50 µm. This corresponds to an acceleration gradient of 0.4 TeV/m. The acceleration length for this study was limited by computational feasibility. However, the structure of the wake and temporal evolution appears similar to that of the two dimensional simulations. It is likely that optimization of parameters in the 3D case could improve the ultimate energy spread of the accelerated beam.
\begin{figure}[H]
\includegraphics[width=0.48\textwidth]{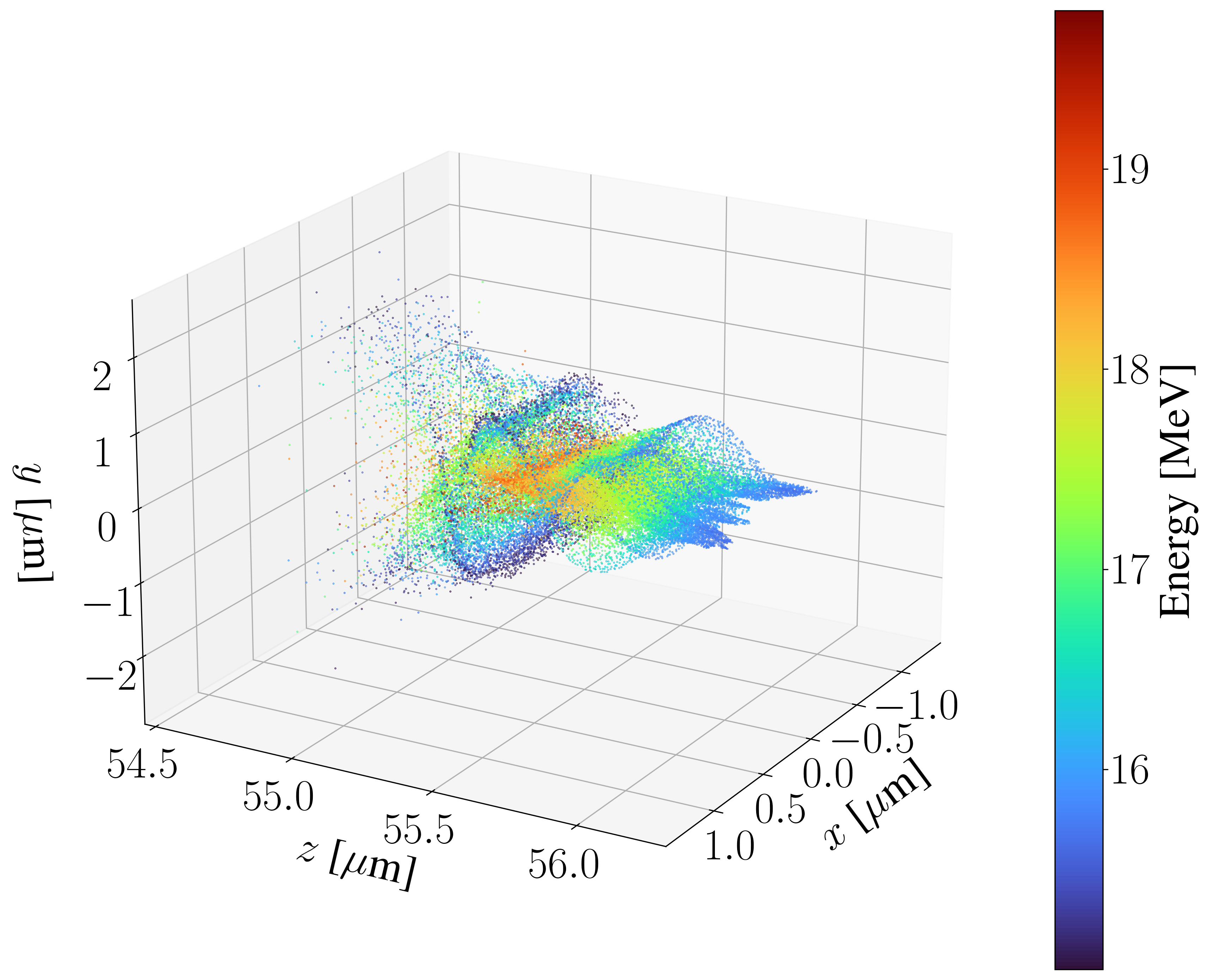}
\caption{\label{Fig:3D_beam} Scatter plot of 3D electron beam. Energy of particles is given in the color bar. Note that this scatter plot shows electrons with energies $>$ 15 MeV.}
\end{figure}
\indent A 3D plot of the electron beam is presented in Figure \ref{Fig:3D_beam} for energies $> 15$ MeV. The beam is slightly elongated in the $\hat{x}$ direction as a result of the asymmetric wake. This demonstrates the production of tailorable electron beam shapes. A ribbon beam such as the one produced has applications in high energy physics \cite{malkin2020development}. It is reasonable to assert that changing the configuration of the incoming beamlets could allow for custom wake shapes, leading to unique electron beam properties.
\\
\indent The second three-dimensional simulation is presented in this work to show that the accelerating structure need not occupy the entire width of the plasma column. The parameters for this case, shown in Figure \ref{Fig:wide_3D}, are as follows: plasma column width $= 40$ µm FWHM, spacing between adjacent beamlets $dx = 1.3$ \textmu m, beam waist of lasers $w_0 = 1.64$ \textmu m, pulse duration $\tau = 20$ fs, pulse energy $E = 830$ µJ, laser wavelength $\lambda = 1$ \textmu m and plasma density $\rho = 1.2 \times 10^{18} \text{cm}^{-3}$ (or 0.012$n_{cr}$). These parameters correspond to laser pulses with normalized vector potentials equal to $0.85$. 
\\
\indent The wake in the second case is composed of a plasma bubble that does not take up the entire width of the plasma column. This is an advantage with regards to the practical implementation of the concept, as this shows that a plasma filament is not necessary to produce an accelerating structure. This result suggests that the accelerating structure could be produced in a macroscopic plasma.

\section{Discussion and Conclusions}
\begin{figure}[H]
\includegraphics[width=0.45\textwidth]{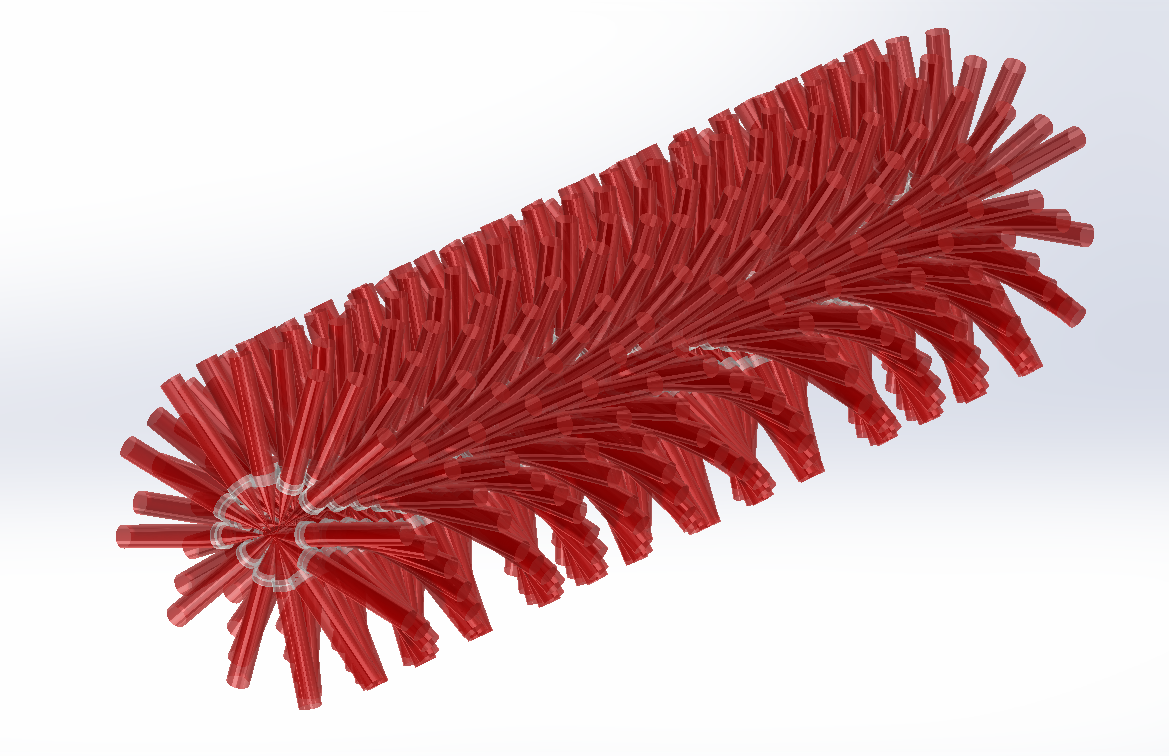}
\caption{\label{Fig:helix} Helical extension to three dimensions. The paths of the beamlets are shown in red and spiral around the central acceleration axis.}
\end{figure}

TPA is a novel acceleration scheme that offers many advantages over existing acceleration concepts. Accelerating ions in an underdense plasma circumvents the need for high contrast laser pulses needed for solid target experiments \cite{kaluza2004influence}. Using multiple moderate intensity laser pulses in tandem with a gas target could also allow for higher repetition rate ion acceleration, something that is presently difficult for solid target experiments. Time-domain multiplexing techniques \cite{zhou2015resonant} provide a means of obtaining intensities/energies needed in the TPA laser pulses. 
\\
\indent 
Traditionally, the three primary limitations to electron acceleration are dephasing, depletion, and defocusing, all of which are bypassed using TPA. Results demonstrated in this work show TPA could allow for the production of a TeV electron beam in $1.9$ m with the potential to be shortened if operated at a higher plasma density.
\\
\indent Perhaps the most unique advantage of TPA is the simplicity of scaling the accelerating structure to longer accelerating distances; scaling only requires additional laser beamlets. There is no need for wavefront shaping of a short pulse or precise chirping of a longer laser pulse, both of which would require specialized optics that may prove difficult to produce. TPA offers a straightforward method of tuning plasma wave phase velocity with applications in ion and electron acceleration.
\\
\indent A potential three dimensional rendering of the TPA concept is shown in Figure \ref{Fig:helix}. Here the paths of the incoming beamlets are shown in red. The focusing optics are gray. Pairs of transversely injected pulses that meet at focus in the middle of the plasma can be arranged in a helix to allow for closer spacing between the adjacent foci. Other arrangements are possible and could offer custom electron beam properties.
\indent TPA could allow for an increase in average power of laser generated particle sources, while simultaneously making them more compact. Low energy lasers can operate at a much higher repetition rate and average power than high energy systems, therefore it is necessary to combine many lower energy pulses to increase flux, while maintaining beam quality.

\begin{acknowledgments}
This work is funded by DOE grant number DE-SC0022109. The authors would like to acknowledge the OSIRIS Consortium, consisting of UCLA and IST (Lisbon, Portugal) for providing access to the OSIRIS 4.0 framework. This research was supported in part through computational resources and services provided by Advanced Research Computing at the University of Michigan, Ann Arbor. 

\end{acknowledgments}
\vspace{15 mm}

\bibliography{Ref}

\providecommand{\noopsort}[1]{}\providecommand{\singleletter}[1]{#1}%
\begin{thebibliography}{17}%
\makeatletter
\providecommand \@ifxundefined [1]{%
 \@ifx{#1\undefined}
}%
\providecommand \@ifnum [1]{%
 \ifnum #1\expandafter \@firstoftwo
 \else \expandafter \@secondoftwo
 \fi
}%
\providecommand \@ifx [1]{%
 \ifx #1\expandafter \@firstoftwo
 \else \expandafter \@secondoftwo
 \fi
}%
\providecommand \natexlab [1]{#1}%
\providecommand \enquote  [1]{``#1''}%
\providecommand \bibnamefont  [1]{#1}%
\providecommand \bibfnamefont [1]{#1}%
\providecommand \citenamefont [1]{#1}%
\providecommand \href@noop [0]{\@secondoftwo}%
\providecommand \href [0]{\begingroup \@sanitize@url \@href}%
\providecommand \@href[1]{\@@startlink{#1}\@@href}%
\providecommand \@@href[1]{\endgroup#1\@@endlink}%
\providecommand \@sanitize@url [0]{\catcode `\\12\catcode `\$12\catcode `\&12\catcode `\#12\catcode `\^12\catcode `\_12\catcode `\%12\relax}%
\providecommand \@@startlink[1]{}%
\providecommand \@@endlink[0]{}%
\providecommand \url  [0]{\begingroup\@sanitize@url \@url }%
\providecommand \@url [1]{\endgroup\@href {#1}{\urlprefix }}%
\providecommand \urlprefix  [0]{URL }%
\providecommand \Eprint [0]{\href }%
\providecommand \doibase [0]{https://doi.org/}%
\providecommand \selectlanguage [0]{\@gobble}%
\providecommand \bibinfo  [0]{\@secondoftwo}%
\providecommand \bibfield  [0]{\@secondoftwo}%
\providecommand \translation [1]{[#1]}%
\providecommand \BibitemOpen [0]{}%
\providecommand \bibitemStop [0]{}%
\providecommand \bibitemNoStop [0]{.\EOS\space}%
\providecommand \EOS [0]{\spacefactor3000\relax}%
\providecommand \BibitemShut  [1]{\csname bibitem#1\endcsname}%
\let\auto@bib@innerbib\@empty
\bibitem [{\citenamefont {Tajima}\ and\ \citenamefont {Dawson}(1979)}]{tajima1979laser}%
  \BibitemOpen
  \bibfield  {author} {\bibinfo {author} {\bibfnamefont {T.}~\bibnamefont {Tajima}}\ and\ \bibinfo {author} {\bibfnamefont {J.~M.}\ \bibnamefont {Dawson}},\ }\bibfield  {title} {\bibinfo {title} {Laser electron accelerator},\ }\href@noop {} {\bibfield  {journal} {\bibinfo  {journal} {Physical Review Letters}\ }\textbf {\bibinfo {volume} {43}},\ \bibinfo {pages} {267} (\bibinfo {year} {1979})}\BibitemShut {NoStop}%
\bibitem [{\citenamefont {Gonsalves}\ \emph {et~al.}(2019)\citenamefont {Gonsalves}, \citenamefont {Nakamura}, \citenamefont {Daniels}, \citenamefont {Benedetti}, \citenamefont {Pieronek}, \citenamefont {De~Raadt}, \citenamefont {Steinke}, \citenamefont {Bin}, \citenamefont {Bulanov}, \citenamefont {Van~Tilborg} \emph {et~al.}}]{gonsalves2019petawatt}%
  \BibitemOpen
  \bibfield  {author} {\bibinfo {author} {\bibfnamefont {A.}~\bibnamefont {Gonsalves}}, \bibinfo {author} {\bibfnamefont {K.}~\bibnamefont {Nakamura}}, \bibinfo {author} {\bibfnamefont {J.}~\bibnamefont {Daniels}}, \bibinfo {author} {\bibfnamefont {C.}~\bibnamefont {Benedetti}}, \bibinfo {author} {\bibfnamefont {C.}~\bibnamefont {Pieronek}}, \bibinfo {author} {\bibfnamefont {T.}~\bibnamefont {De~Raadt}}, \bibinfo {author} {\bibfnamefont {S.}~\bibnamefont {Steinke}}, \bibinfo {author} {\bibfnamefont {J.}~\bibnamefont {Bin}}, \bibinfo {author} {\bibfnamefont {S.}~\bibnamefont {Bulanov}}, \bibinfo {author} {\bibfnamefont {J.}~\bibnamefont {Van~Tilborg}}, \emph {et~al.},\ }\bibfield  {title} {\bibinfo {title} {Petawatt laser guiding and electron beam acceleration to 8 gev in a laser-heated capillary discharge waveguide},\ }\href@noop {} {\bibfield  {journal} {\bibinfo  {journal} {Physical Review Letters}\ }\textbf {\bibinfo {volume} {122}},\ \bibinfo {pages} {084801} (\bibinfo {year} {2019})}\BibitemShut
  {NoStop}%
\bibitem [{\citenamefont {Aniculaesei}\ \emph {et~al.}(2024)\citenamefont {Aniculaesei}, \citenamefont {Ha}, \citenamefont {Yoffe}, \citenamefont {Labun}, \citenamefont {Milton}, \citenamefont {McCary}, \citenamefont {Spinks}, \citenamefont {Quevedo}, \citenamefont {Labun}, \citenamefont {Sain} \emph {et~al.}}]{aniculaesei2024acceleration}%
  \BibitemOpen
  \bibfield  {author} {\bibinfo {author} {\bibfnamefont {C.}~\bibnamefont {Aniculaesei}}, \bibinfo {author} {\bibfnamefont {T.}~\bibnamefont {Ha}}, \bibinfo {author} {\bibfnamefont {S.}~\bibnamefont {Yoffe}}, \bibinfo {author} {\bibfnamefont {L.}~\bibnamefont {Labun}}, \bibinfo {author} {\bibfnamefont {S.}~\bibnamefont {Milton}}, \bibinfo {author} {\bibfnamefont {E.}~\bibnamefont {McCary}}, \bibinfo {author} {\bibfnamefont {M.~M.}\ \bibnamefont {Spinks}}, \bibinfo {author} {\bibfnamefont {H.~J.}\ \bibnamefont {Quevedo}}, \bibinfo {author} {\bibfnamefont {O.~Z.}\ \bibnamefont {Labun}}, \bibinfo {author} {\bibfnamefont {R.}~\bibnamefont {Sain}}, \emph {et~al.},\ }\bibfield  {title} {\bibinfo {title} {The acceleration of a high-charge electron bunch to 10 gev in a 10-cm nanoparticle-assisted wakefield accelerator},\ }\href@noop {} {\bibfield  {journal} {\bibinfo  {journal} {Matter and Radiation at Extremes}\ }\textbf {\bibinfo {volume} {9}} (\bibinfo {year} {2024})}\BibitemShut {NoStop}%
\bibitem [{\citenamefont {Froula}\ \emph {et~al.}(2018)\citenamefont {Froula}, \citenamefont {Turnbull}, \citenamefont {Davies}, \citenamefont {Kessler}, \citenamefont {Haberberger}, \citenamefont {Palastro}, \citenamefont {Bahk}, \citenamefont {Begishev}, \citenamefont {Boni}, \citenamefont {Bucht} \emph {et~al.}}]{froula2018spatiotemporal}%
  \BibitemOpen
  \bibfield  {author} {\bibinfo {author} {\bibfnamefont {D.~H.}\ \bibnamefont {Froula}}, \bibinfo {author} {\bibfnamefont {D.}~\bibnamefont {Turnbull}}, \bibinfo {author} {\bibfnamefont {A.~S.}\ \bibnamefont {Davies}}, \bibinfo {author} {\bibfnamefont {T.~J.}\ \bibnamefont {Kessler}}, \bibinfo {author} {\bibfnamefont {D.}~\bibnamefont {Haberberger}}, \bibinfo {author} {\bibfnamefont {J.~P.}\ \bibnamefont {Palastro}}, \bibinfo {author} {\bibfnamefont {S.-W.}\ \bibnamefont {Bahk}}, \bibinfo {author} {\bibfnamefont {I.~A.}\ \bibnamefont {Begishev}}, \bibinfo {author} {\bibfnamefont {R.}~\bibnamefont {Boni}}, \bibinfo {author} {\bibfnamefont {S.}~\bibnamefont {Bucht}}, \emph {et~al.},\ }\bibfield  {title} {\bibinfo {title} {Spatiotemporal control of laser intensity},\ }\href@noop {} {\bibfield  {journal} {\bibinfo  {journal} {Nature Photonics}\ }\textbf {\bibinfo {volume} {12}},\ \bibinfo {pages} {262} (\bibinfo {year} {2018})}\BibitemShut {NoStop}%
\bibitem [{\citenamefont {Froula}\ \emph {et~al.}(2019)\citenamefont {Froula}, \citenamefont {Palastro}, \citenamefont {Turnbull}, \citenamefont {Davies}, \citenamefont {Nguyen}, \citenamefont {Howard}, \citenamefont {Ramsey}, \citenamefont {Franke}, \citenamefont {Bahk}, \citenamefont {Begishev} \emph {et~al.}}]{froula2019flying}%
  \BibitemOpen
  \bibfield  {author} {\bibinfo {author} {\bibfnamefont {D.}~\bibnamefont {Froula}}, \bibinfo {author} {\bibfnamefont {J.}~\bibnamefont {Palastro}}, \bibinfo {author} {\bibfnamefont {D.}~\bibnamefont {Turnbull}}, \bibinfo {author} {\bibfnamefont {A.}~\bibnamefont {Davies}}, \bibinfo {author} {\bibfnamefont {L.}~\bibnamefont {Nguyen}}, \bibinfo {author} {\bibfnamefont {A.}~\bibnamefont {Howard}}, \bibinfo {author} {\bibfnamefont {D.}~\bibnamefont {Ramsey}}, \bibinfo {author} {\bibfnamefont {P.}~\bibnamefont {Franke}}, \bibinfo {author} {\bibfnamefont {S.-W.}\ \bibnamefont {Bahk}}, \bibinfo {author} {\bibfnamefont {I.}~\bibnamefont {Begishev}}, \emph {et~al.},\ }\bibfield  {title} {\bibinfo {title} {Flying focus: Spatial and temporal control of intensity for laser-based applications},\ }\href@noop {} {\bibfield  {journal} {\bibinfo  {journal} {Physics of Plasmas}\ }\textbf {\bibinfo {volume} {26}} (\bibinfo {year} {2019})}\BibitemShut {NoStop}%
\bibitem [{\citenamefont {Miller}\ \emph {et~al.}(2023)\citenamefont {Miller}, \citenamefont {Pierce}, \citenamefont {Ambat}, \citenamefont {Shaw}, \citenamefont {Weichman}, \citenamefont {Mori}, \citenamefont {Froula},\ and\ \citenamefont {Palastro}}]{miller2023dephasingless}%
  \BibitemOpen
  \bibfield  {author} {\bibinfo {author} {\bibfnamefont {K.~G.}\ \bibnamefont {Miller}}, \bibinfo {author} {\bibfnamefont {J.~R.}\ \bibnamefont {Pierce}}, \bibinfo {author} {\bibfnamefont {M.~V.}\ \bibnamefont {Ambat}}, \bibinfo {author} {\bibfnamefont {J.~L.}\ \bibnamefont {Shaw}}, \bibinfo {author} {\bibfnamefont {K.}~\bibnamefont {Weichman}}, \bibinfo {author} {\bibfnamefont {W.~B.}\ \bibnamefont {Mori}}, \bibinfo {author} {\bibfnamefont {D.~H.}\ \bibnamefont {Froula}},\ and\ \bibinfo {author} {\bibfnamefont {J.~P.}\ \bibnamefont {Palastro}},\ }\bibfield  {title} {\bibinfo {title} {Dephasingless laser wakefield acceleration in the bubble regime},\ }\href@noop {} {\bibfield  {journal} {\bibinfo  {journal} {Scientific Reports}\ }\textbf {\bibinfo {volume} {13}},\ \bibinfo {pages} {21306} (\bibinfo {year} {2023})}\BibitemShut {NoStop}%
\bibitem [{\citenamefont {Pierce}\ \emph {et~al.}(2023)\citenamefont {Pierce}, \citenamefont {Palastro}, \citenamefont {Li}, \citenamefont {Malaca}, \citenamefont {Ramsey}, \citenamefont {Vieira}, \citenamefont {Weichman},\ and\ \citenamefont {Mori}}]{pierce2023arbitrarily}%
  \BibitemOpen
  \bibfield  {author} {\bibinfo {author} {\bibfnamefont {J.~R.}\ \bibnamefont {Pierce}}, \bibinfo {author} {\bibfnamefont {J.~P.}\ \bibnamefont {Palastro}}, \bibinfo {author} {\bibfnamefont {F.}~\bibnamefont {Li}}, \bibinfo {author} {\bibfnamefont {B.}~\bibnamefont {Malaca}}, \bibinfo {author} {\bibfnamefont {D.}~\bibnamefont {Ramsey}}, \bibinfo {author} {\bibfnamefont {J.}~\bibnamefont {Vieira}}, \bibinfo {author} {\bibfnamefont {K.}~\bibnamefont {Weichman}},\ and\ \bibinfo {author} {\bibfnamefont {W.~B.}\ \bibnamefont {Mori}},\ }\bibfield  {title} {\bibinfo {title} {Arbitrarily structured laser pulses},\ }\href@noop {} {\bibfield  {journal} {\bibinfo  {journal} {Physical Review Research}\ }\textbf {\bibinfo {volume} {5}},\ \bibinfo {pages} {013085} (\bibinfo {year} {2023})}\BibitemShut {NoStop}%
\bibitem [{\citenamefont {Palastro}\ \emph {et~al.}(2020)\citenamefont {Palastro}, \citenamefont {Shaw}, \citenamefont {Franke}, \citenamefont {Ramsey}, \citenamefont {Simpson},\ and\ \citenamefont {Froula}}]{palastro2020dephasingless}%
  \BibitemOpen
  \bibfield  {author} {\bibinfo {author} {\bibfnamefont {J.}~\bibnamefont {Palastro}}, \bibinfo {author} {\bibfnamefont {J.}~\bibnamefont {Shaw}}, \bibinfo {author} {\bibfnamefont {P.}~\bibnamefont {Franke}}, \bibinfo {author} {\bibfnamefont {D.}~\bibnamefont {Ramsey}}, \bibinfo {author} {\bibfnamefont {T.}~\bibnamefont {Simpson}},\ and\ \bibinfo {author} {\bibfnamefont {D.}~\bibnamefont {Froula}},\ }\bibfield  {title} {\bibinfo {title} {Dephasingless laser wakefield acceleration},\ }\href@noop {} {\bibfield  {journal} {\bibinfo  {journal} {Physical Review Letters}\ }\textbf {\bibinfo {volume} {124}},\ \bibinfo {pages} {134802} (\bibinfo {year} {2020})}\BibitemShut {NoStop}%
\bibitem [{\citenamefont {Gong}\ \emph {et~al.}(2024)\citenamefont {Gong}, \citenamefont {Cao}, \citenamefont {Palastro},\ and\ \citenamefont {Edwards}}]{gong2024laser}%
  \BibitemOpen
  \bibfield  {author} {\bibinfo {author} {\bibfnamefont {Z.}~\bibnamefont {Gong}}, \bibinfo {author} {\bibfnamefont {S.}~\bibnamefont {Cao}}, \bibinfo {author} {\bibfnamefont {J.~P.}\ \bibnamefont {Palastro}},\ and\ \bibinfo {author} {\bibfnamefont {M.~R.}\ \bibnamefont {Edwards}},\ }\bibfield  {title} {\bibinfo {title} {Laser wakefield acceleration of ions with a transverse flying focus},\ }\href {https://doi.org/10.1103/PhysRevLett.133.265002} {\bibfield  {journal} {\bibinfo  {journal} {Phys. Rev. Lett.}\ }\textbf {\bibinfo {volume} {133}},\ \bibinfo {pages} {265002} (\bibinfo {year} {2024})}\BibitemShut {NoStop}%
\bibitem [{\citenamefont {Hinchliffe}\ and\ \citenamefont {Battaglia}(2004)}]{hinchliffe2004tev}%
  \BibitemOpen
  \bibfield  {author} {\bibinfo {author} {\bibfnamefont {I.}~\bibnamefont {Hinchliffe}}\ and\ \bibinfo {author} {\bibfnamefont {M.}~\bibnamefont {Battaglia}},\ }\bibfield  {title} {\bibinfo {title} {A tev linear collider},\ }\href@noop {} {\bibfield  {journal} {\bibinfo  {journal} {Physics Today}\ }\textbf {\bibinfo {volume} {57}},\ \bibinfo {pages} {49} (\bibinfo {year} {2004})}\BibitemShut {NoStop}%
\bibitem [{\citenamefont {Debus}\ \emph {et~al.}(2019)\citenamefont {Debus}, \citenamefont {Pausch}, \citenamefont {Huebl}, \citenamefont {Steiniger}, \citenamefont {Widera}, \citenamefont {Cowan}, \citenamefont {Schramm},\ and\ \citenamefont {Bussmann}}]{debus2019circumventing}%
  \BibitemOpen
  \bibfield  {author} {\bibinfo {author} {\bibfnamefont {A.}~\bibnamefont {Debus}}, \bibinfo {author} {\bibfnamefont {R.}~\bibnamefont {Pausch}}, \bibinfo {author} {\bibfnamefont {A.}~\bibnamefont {Huebl}}, \bibinfo {author} {\bibfnamefont {K.}~\bibnamefont {Steiniger}}, \bibinfo {author} {\bibfnamefont {R.}~\bibnamefont {Widera}}, \bibinfo {author} {\bibfnamefont {T.~E.}\ \bibnamefont {Cowan}}, \bibinfo {author} {\bibfnamefont {U.}~\bibnamefont {Schramm}},\ and\ \bibinfo {author} {\bibfnamefont {M.}~\bibnamefont {Bussmann}},\ }\bibfield  {title} {\bibinfo {title} {Circumventing the dephasing and depletion limits of laser-wakefield acceleration},\ }\href@noop {} {\bibfield  {journal} {\bibinfo  {journal} {Physical Review X}\ }\textbf {\bibinfo {volume} {9}},\ \bibinfo {pages} {031044} (\bibinfo {year} {2019})}\BibitemShut {NoStop}%
\bibitem [{\citenamefont {Macchi}\ \emph {et~al.}(2010)\citenamefont {Macchi}, \citenamefont {Veghini}, \citenamefont {Liseykina},\ and\ \citenamefont {Pegoraro}}]{macchi2010radiation}%
  \BibitemOpen
  \bibfield  {author} {\bibinfo {author} {\bibfnamefont {A.}~\bibnamefont {Macchi}}, \bibinfo {author} {\bibfnamefont {S.}~\bibnamefont {Veghini}}, \bibinfo {author} {\bibfnamefont {T.~V.}\ \bibnamefont {Liseykina}},\ and\ \bibinfo {author} {\bibfnamefont {F.}~\bibnamefont {Pegoraro}},\ }\bibfield  {title} {\bibinfo {title} {Radiation pressure acceleration of ultrathin foils},\ }\href@noop {} {\bibfield  {journal} {\bibinfo  {journal} {New Journal of Physics}\ }\textbf {\bibinfo {volume} {12}},\ \bibinfo {pages} {045013} (\bibinfo {year} {2010})}\BibitemShut {NoStop}%
\bibitem [{\citenamefont {Robinson}\ \emph {et~al.}(2008)\citenamefont {Robinson}, \citenamefont {Zepf}, \citenamefont {Kar}, \citenamefont {Evans},\ and\ \citenamefont {Bellei}}]{robinson2008radiation}%
  \BibitemOpen
  \bibfield  {author} {\bibinfo {author} {\bibfnamefont {A.}~\bibnamefont {Robinson}}, \bibinfo {author} {\bibfnamefont {M.}~\bibnamefont {Zepf}}, \bibinfo {author} {\bibfnamefont {S.}~\bibnamefont {Kar}}, \bibinfo {author} {\bibfnamefont {R.}~\bibnamefont {Evans}},\ and\ \bibinfo {author} {\bibfnamefont {C.}~\bibnamefont {Bellei}},\ }\bibfield  {title} {\bibinfo {title} {Radiation pressure acceleration of thin foils with circularly polarized laser pulses},\ }\href@noop {} {\bibfield  {journal} {\bibinfo  {journal} {New Journal of Physics}\ }\textbf {\bibinfo {volume} {10}},\ \bibinfo {pages} {013021} (\bibinfo {year} {2008})}\BibitemShut {NoStop}%
\bibitem [{\citenamefont {Lu}\ \emph {et~al.}(2007)\citenamefont {Lu}, \citenamefont {Tzoufras}, \citenamefont {Joshi}, \citenamefont {Tsung}, \citenamefont {Mori}, \citenamefont {Vieira}, \citenamefont {Fonseca},\ and\ \citenamefont {Silva}}]{PhysRevSTAB.10.061301}%
  \BibitemOpen
  \bibfield  {author} {\bibinfo {author} {\bibfnamefont {W.}~\bibnamefont {Lu}}, \bibinfo {author} {\bibfnamefont {M.}~\bibnamefont {Tzoufras}}, \bibinfo {author} {\bibfnamefont {C.}~\bibnamefont {Joshi}}, \bibinfo {author} {\bibfnamefont {F.~S.}\ \bibnamefont {Tsung}}, \bibinfo {author} {\bibfnamefont {W.~B.}\ \bibnamefont {Mori}}, \bibinfo {author} {\bibfnamefont {J.}~\bibnamefont {Vieira}}, \bibinfo {author} {\bibfnamefont {R.~A.}\ \bibnamefont {Fonseca}},\ and\ \bibinfo {author} {\bibfnamefont {L.~O.}\ \bibnamefont {Silva}},\ }\bibfield  {title} {\bibinfo {title} {Generating multi-gev electron bunches using single stage laser wakefield acceleration in a 3d nonlinear regime},\ }\href {https://doi.org/10.1103/PhysRevSTAB.10.061301} {\bibfield  {journal} {\bibinfo  {journal} {Phys. Rev. ST Accel. Beams}\ }\textbf {\bibinfo {volume} {10}},\ \bibinfo {pages} {061301} (\bibinfo {year} {2007})}\BibitemShut {NoStop}%
\bibitem [{\citenamefont {Malkin}\ \emph {et~al.}(2020)\citenamefont {Malkin}, \citenamefont {Zaslavsky}, \citenamefont {Zheleznov}, \citenamefont {Goykhman}, \citenamefont {Gromov}, \citenamefont {Palitsin}, \citenamefont {Sergeev}, \citenamefont {Fedotov}, \citenamefont {Makhalov},\ and\ \citenamefont {Ginzburg}}]{malkin2020development}%
  \BibitemOpen
  \bibfield  {author} {\bibinfo {author} {\bibfnamefont {A.}~\bibnamefont {Malkin}}, \bibinfo {author} {\bibfnamefont {V.~Y.}\ \bibnamefont {Zaslavsky}}, \bibinfo {author} {\bibfnamefont {I.}~\bibnamefont {Zheleznov}}, \bibinfo {author} {\bibfnamefont {M.}~\bibnamefont {Goykhman}}, \bibinfo {author} {\bibfnamefont {A.}~\bibnamefont {Gromov}}, \bibinfo {author} {\bibfnamefont {A.}~\bibnamefont {Palitsin}}, \bibinfo {author} {\bibfnamefont {A.}~\bibnamefont {Sergeev}}, \bibinfo {author} {\bibfnamefont {A.}~\bibnamefont {Fedotov}}, \bibinfo {author} {\bibfnamefont {P.}~\bibnamefont {Makhalov}},\ and\ \bibinfo {author} {\bibfnamefont {N.}~\bibnamefont {Ginzburg}},\ }\bibfield  {title} {\bibinfo {title} {Development of high-power millimeter-wave surface-wave generators based on relativistic ribbon electron beams},\ }\href@noop {} {\bibfield  {journal} {\bibinfo  {journal} {Radiophysics and quantum electronics}\ }\textbf {\bibinfo {volume} {63}},\ \bibinfo {pages} {458} (\bibinfo {year} {2020})}\BibitemShut
  {NoStop}%
\bibitem [{\citenamefont {Kaluza}\ \emph {et~al.}(2004)\citenamefont {Kaluza}, \citenamefont {Schreiber}, \citenamefont {Santala}, \citenamefont {Tsakiris}, \citenamefont {Eidmann}, \citenamefont {Meyer-ter Vehn},\ and\ \citenamefont {Witte}}]{kaluza2004influence}%
  \BibitemOpen
  \bibfield  {author} {\bibinfo {author} {\bibfnamefont {M.}~\bibnamefont {Kaluza}}, \bibinfo {author} {\bibfnamefont {J.}~\bibnamefont {Schreiber}}, \bibinfo {author} {\bibfnamefont {M.~I.}\ \bibnamefont {Santala}}, \bibinfo {author} {\bibfnamefont {G.~D.}\ \bibnamefont {Tsakiris}}, \bibinfo {author} {\bibfnamefont {K.}~\bibnamefont {Eidmann}}, \bibinfo {author} {\bibfnamefont {J.}~\bibnamefont {Meyer-ter Vehn}},\ and\ \bibinfo {author} {\bibfnamefont {K.~J.}\ \bibnamefont {Witte}},\ }\bibfield  {title} {\bibinfo {title} {Influence of the laser prepulse on proton acceleration in thin-foil experiments},\ }\href@noop {} {\bibfield  {journal} {\bibinfo  {journal} {Physical Review Letters}\ }\textbf {\bibinfo {volume} {93}},\ \bibinfo {pages} {045003} (\bibinfo {year} {2004})}\BibitemShut {NoStop}%
\bibitem [{\citenamefont {Zhou}\ \emph {et~al.}(2015)\citenamefont {Zhou}, \citenamefont {Ruppe}, \citenamefont {Stanfield}, \citenamefont {Nees}, \citenamefont {Wilcox},\ and\ \citenamefont {Galvanauskas}}]{zhou2015resonant}%
  \BibitemOpen
  \bibfield  {author} {\bibinfo {author} {\bibfnamefont {T.}~\bibnamefont {Zhou}}, \bibinfo {author} {\bibfnamefont {J.}~\bibnamefont {Ruppe}}, \bibinfo {author} {\bibfnamefont {P.}~\bibnamefont {Stanfield}}, \bibinfo {author} {\bibfnamefont {J.}~\bibnamefont {Nees}}, \bibinfo {author} {\bibfnamefont {R.}~\bibnamefont {Wilcox}},\ and\ \bibinfo {author} {\bibfnamefont {A.}~\bibnamefont {Galvanauskas}},\ }\bibfield  {title} {\bibinfo {title} {Resonant cavity based time-domain multiplexing techniques for coherently combined fiber laser systems},\ }\href@noop {} {\bibfield  {journal} {\bibinfo  {journal} {The European Physical Journal Special Topics}\ }\textbf {\bibinfo {volume} {224}},\ \bibinfo {pages} {2585} (\bibinfo {year} {2015})}\BibitemShut {NoStop}%
\end{thebibliography}%

\end{document}